\documentclass[preprint,authoryear]{elsarticle}

\usepackage{amssymb}
\usepackage{amsmath}
\usepackage{array}
\usepackage{tabularx}
\usepackage{booktabs}
\usepackage{graphicx}
\usepackage{hyperref}
\usepackage{tikz}
\usetikzlibrary{positioning,shapes.geometric,arrows.meta,fit,calc,decorations.pathreplacing}
\date{}
\makeatletter
\def\ps@pprintTitle{
  \let\@oddhead\@empty
  \let\@evenhead\@empty
  \def\@oddfoot{}
  \let\@evenfoot\@oddfoot
}
\makeatother

\newcolumntype{L}[1]{>{\raggedright\arraybackslash}p{#1}}
\newcommand{\code}[1]{\nolinkurl{#1}}
\newcommand{\filepath}[1]{\path{#1}}

\begin{document}
\begin{frontmatter}

\title{CHSR-RRF: A curriculum-gated hybrid retrieval framework with reciprocal rank fusion and leakage-aware benchmarking for educational RAG}

\author[uco]{Terence Ateya}
\author[tamu]{Zavier Ndum Ndum}
\author[uco]{Jicheng Fu}
\author[ou]{Kelly Tendongkeng}

\affiliation[uco]{organization={University of Central Oklahoma},
            addressline={Department of Computer Science and Mathematics},
            city={Edmond},
            state={Oklahoma},
            country={United States}}

\affiliation[tamu]{organization={Texas A\&M University},
            addressline={Department of Nuclear Engineering},
            city={College Station},
            state={Texas},
            country={United States}}

\affiliation[ou]{organization={University of Oklahoma},
            addressline={Gallogly College of Engineering, School of Data Science and Analytics},
            city={Norman},
            state={Oklahoma},
            country={United States}}
\begin{abstract}
Retrieval-augmented generation (RAG) is increasingly used in educational question answering, but standard retrievers optimize topical relevance without enforcing curriculum validity. In school settings, a passage can be relevant yet inappropriate if it comes from the wrong subject, level, or examination context; we call this failure mode curriculum leakage. We present CHSR-RRF, a curriculum-gated hybrid retrieval framework that applies metadata constraints before retrieval, then combines sparse and dense search with reciprocal rank fusion and deterministic reranking. We also introduce CERB, a 126-case benchmark for curriculum-constrained retrieval with hierarchy-aware relevance labels and explicit leakage annotations. On a 61-case pilot, pre-retrieval gating reduces leakage by 4.6x ($p<0.001$) while preserving ranked recall, whereas applying the same constraints after retrieval collapses recall and exact-scope success to zero ($p=0.039$). A full-benchmark lower-bound analysis further shows that many remaining failures arise from corpus and metadata gaps rather than retrieval design alone. These results show that retrieval in structured educational domains should be treated as constrained selection, with validity enforced when the candidate pool is formed rather than after ranking.
\end{abstract}


\end{frontmatter}

\begin{table}[htbp]
\centering
\small
\caption{Notation and abbreviations used throughout the paper.}
\label{tab:notation}
\begin{tabularx}{\linewidth}{@{}lX@{}}
\toprule
\textbf{Symbol / abbreviation} & \textbf{Definition} \\
\midrule
\multicolumn{2}{@{}l}{\emph{Abbreviations}}\\
RAG & Retrieval-augmented generation \\
LLM & Large language model \\
CHSR-RRF & Curriculum-Hard-gated Scoped Retrieval with Reciprocal Rank Fusion (this paper) \\
CERB & Cameroon Exam Retrieval Benchmark (this paper) \\
BM25 & Best Match 25 (lexical ranking function) \\
FTS & Full-text search (lexical retrieval) \\
RRF & Reciprocal rank fusion \\
nDCG & Normalized discounted cumulative gain \\
ESS & Exact-scope success (at least one returned document matches all expected metadata) \\
HNSW & Hierarchical Navigable Small World (approximate nearest-neighbour index) \\
DPR & Dense passage retrieval \\
IAA & Inter-annotator agreement \\
MCQ & Multiple-choice question \\
GCE & (Cameroon) General Certificate of Education \\
O-Level / A-Level & Ordinary Level / Advanced Level \\
\midrule
\multicolumn{2}{@{}l}{\emph{Mathematical notation}}\\
$D$ & Corpus of educational evidence chunks \\
$d \in D$ & A single evidence chunk \\
$q$ & User query (natural language) \\
$h(q) = (s, \ell, p, t, \tau)$ & Extracted hierarchy: subject $s$, level $\ell$, paper $p$, topic $t$, subtopic $\tau$ \\
$c(d) = (s_d, \ell_d, p_d, t_d, \tau_d)$ & Curriculum metadata of chunk $d$ \\
$G(d, q)$ & Hard curriculum gate (binary) \\
$D_g(q)$ & Gated candidate pool: $\{d \in D \mid G(d, q) = 1\}$ \\
$\mathbf{e}(\cdot)$ & Embedding function (text-embedding-3-small, 1536-dim) \\
$\kappa$ & RRF constant (default 60) \\
$k_{\min}$ & Minimum result count before widening triggers \\
$W_i$ & Widening level $i$ (progressive constraint relaxation) \\
\bottomrule
\end{tabularx}
\end{table}

\section{Introduction}
\label{sec:intro}

Retrieval-Augmented Generation (RAG) has become a standard way to ground large language models (LLMs) in outside knowledge \citep{Lewis2020RAG, Gao2023RAGSurvey, Fan2024RAGSurvey}. Most systems are tuned for semantic relevance: they assume that finding topically similar documents is enough for a correct answer. In structured domains such as education, that assumption breaks down.

In exam-oriented education, correctness is not only about topic. It is also about \emph{curriculum admissibility}: the retrieved document must fall within the valid scope of the learner's subject, level, and examination context. A document can be on-topic and still be wrong for the learner if it comes from the wrong subject, level, examination paper, or session. For example, returning A-Level physics material for an O-Level query introduces concepts the student has not met (relativistic momentum, center-of-mass frames), which confuses rather than helps. We refer to this failure mode as \emph{curriculum leakage}: the retrieval of documents that break the curriculum scope implied by the query, even when those documents are on-topic. This is not an edge case. In the 61-case pilot evaluation reported here (drawn from the 126-case CERB benchmark; see Section~\ref{par:pilot_subset}), unconstrained hybrid retrieval has a leakage rate of \mbox{0.7596}: more than three-quarters of retrieved documents come from the wrong curriculum scope.

This exposes a structural limit in current RAG systems on structured educational corpora. They treat retrieval as an unconstrained ranking problem. It should instead be a \emph{constrained selection problem}, where curriculum admissibility is enforced when the candidate pool is built. Prior work improves retrieval through hybrid search \citep{Ma2021CLEAR, Cormack2009RRF}, iterative reasoning \citep{Trivedi2023IRCoT}, or self-reflection \citep{Asai2024SelfRAG}, but all of it ranks over an unconstrained candidate space and does not enforce structured constraints before retrieval. Even the best curriculum-gated variant in our study has a leakage rate of \mbox{0.1653} and an exact-scope success (ESS, the share of queries for which at least one retrieved document matches all expected curriculum metadata fields) of only \mbox{0.0492}. Applying metadata filters after retrieval rather than before is worse still: it collapses recall to zero, because it discards most results without finding valid replacements. Prompting the generator to ``stay on topic'' does not fix this, because the error happens before generation.

We therefore argue that educational retrieval should be reframed as a \emph{constrained retrieval problem}, where relevance is optimised subject to curriculum-validity constraints. This matters most for high-accountability AI systems, such as national examination tutoring, medical education, and legal training, where out-of-scope information harms learning. The goal of this paper is not to maximise benchmark scores. It is to expose a failure mode no one has measured before and to give a framework for studying it. We show that systems tuned for semantic relevance alone can look effective under standard evaluation while routinely retrieving material from the wrong subject or level. By making admissibility explicit and measurable, we move the evaluation target from ``what is relevant'' to ``what is valid within scope.'' The retrieval layer acts as the admissibility gate for any downstream agentic system: if the retriever admits invalid evidence, no amount of later reasoning can recover curriculum safety.

Table~\ref{tab:summary} previews the whole study at a glance: the corpus, the two evaluation phases, the variants compared, the metrics, and the headline result.

\begin{table}[htbp]
\centering
\small
\caption{Study at a glance: corpus, evaluation phases, variants, metrics, and headline result.}
\label{tab:summary}
\setlength{\tabcolsep}{4pt}
\resizebox{\linewidth}{!}{%
\begin{tabular}{@{}L{0.20\linewidth}L{0.80\linewidth}@{}}
\toprule
\textbf{Element} & \textbf{Summary} \\
\midrule
Corpus & 74{,}018 indexed evidence chunks from $>$2{,}000 GCE source documents across 11 subjects; frozen evaluation snapshot. \\
Benchmark & CERB: 126 cases (one student query each) across 7 task slices, with hierarchy-aware relevance, leakage, and exact-scope labels. \\
Phase 1 (pilot) & Six-variant paired-bootstrap ablation ($N{=}1000$, seed=42) on a 61-case pilot subset (Section~\ref{sec:results}). \\
Phase 2 (final corpus) & Single-variant \code{fts_only_scoped} measurement and a paired within-corpus probe on the full 126 cases (Section~\ref{sec:postfix_validation}). \\
Variants & A dense-only, B hybrid ungated, B+ hybrid post-filtered, C hybrid gated, D gated+reranked, E full CHSR-RRF. \\
Metrics & Precision@8, Recall@8, nDCG@8, leakage rate, contamination rate, exact-scope success (ESS), latency. \\
Headline & Gating \emph{before} search cuts leakage $4.6\times$ ($0.7596\!\to\!0.1653$, $p<0.001$) with recall preserved; the same filters applied \emph{after} search collapse recall and ESS to zero. \\
\bottomrule
\end{tabular}
}
\end{table}

\subsection{Research Questions}

This work is guided by three core research questions:

\begin{enumerate}
    \item \textbf{RQ1:} Does enforcing curriculum constraints improve retrieval reliability compared to unconstrained semantic retrieval in educational RAG systems?
    \item \textbf{RQ2:} What is the trade-off between retrieval quality and curriculum leakage when constraints are relaxed through controlled widening?
    \item \textbf{RQ3:} Can curriculum-aware retrieval strategies improve exact-scope evidence recovery in structured educational benchmarks?
\end{enumerate}

These questions frame retrieval not only as a ranking problem, but as a constrained decision process where systems must balance relevance, admissibility, and coverage.

To answer them, we instantiate curriculum-constrained retrieval as CHSR-RRF, a retrieval framework that treats curriculum structure as a first-class constraint and enforces admissibility \emph{before} ranking. The central empirical result, established on a 61-case pilot subset of the 126-case CERB benchmark (Section~\ref{par:pilot_subset}), is a structural asymmetry between enforcement points: gating before retrieval reduces leakage by $4.6\times$ ($p < 0.001$) while keeping recall, whereas applying the same metadata filters after retrieval collapses recall and exact-scope success to zero ($p = 0.039$ on recall for the $\text{B+} \!\to\! \text{C}$ contrast). A progressive, auditable scope-relaxation policy, used only when the gated pipeline returns too few results, recovers limited cross-subject recall at a measurable leakage and latency cost, which we report as a secondary trade-off. CERB and the per-slice root-cause analysis are the artefacts that make these claims measurable.

\subsection{Contributions}

The primary contribution of this paper is the empirical demonstration of \emph{when} and \emph{why} pre-retrieval curriculum gating cannot be replaced by post-hoc filtering in structured educational retrieval. In other words, the headline is a finding, and CERB and the per-slice root-cause analysis are the two artefacts that make the finding measurable.

\begin{enumerate}
    \item \textbf{Primary: the finding (mechanism).} We isolate pre-retrieval curriculum gating as the load-bearing mechanism: applying the same metadata constraints before retrieval (Variant~C) versus after retrieval (Variant~B+) produces qualitatively different outcomes ($p = 0.039$ on recall, $p < 0.001$ on leakage). This is not a ranking improvement but a candidate-pool composition effect that post-hoc correction cannot recover. We instantiate it in CHSR-RRF, a five-stage deterministic pipeline.
    \item \textbf{Artefact 1: the benchmark.} We release CERB (Cameroon Exam Retrieval Benchmark), a 126-case benchmark protocol with hierarchy-aware relevance labels, explicit curriculum-leakage annotations, and seven task-specific slices. The protocol is system-agnostic; the Cameroon GCE instantiation is the first concrete instance.
    \item \textbf{Artefact 2: the ablation.} A six-variant paired-bootstrap ablation ($N{=}1000$, seed=42) on a 61-case pilot subset of CERB establishes the gating-versus-filtering asymmetry above and ranks the marginal contribution of every other pipeline stage, including a $4.6\times$ leakage reduction from gating alone (ungated hybrid B $\to$ gated hybrid C, $0.7596 \to 0.1653$, $p < 0.001$).
    \item \textbf{Supporting: the diagnostic.} We provide a per-slice root-cause analysis over the five lowest-performing CERB slices (36 cases total) that separates data gaps (21 cases), preprocessing gaps (10 cases), and method limitations (5 cases), and validate it with a post-fix \texttt{fts\_only} measurement on the full 126-case benchmark (\S\ref{sec:postfix_validation}): preprocessing-classified slices recover (formula-heavy 0.050~$\to$~0.222; bilingual 0.000~$\to$~0.100), while corpus-coverage-classified slices remain at zero as predicted (mark-scheme, table-heavy).
\end{enumerate}

\section{Related Work and Background}
\label{sec:related}

\subsection{Sparse, Dense, and Hybrid Retrieval}

Lexical ranking with BM25 (Best Match 25) and its probabilistic variants has anchored information retrieval for over two decades (Robertson and Zaragoza, 2009). BM25 scores a query against a document by term overlap; it remains a strong baseline and is still the default first stage in many production stacks. Dense retrievers replaced term overlap with learned vectors: Karpukhin and colleagues introduced dense passage retrieval (DPR), which learns query and document embeddings from supervised question--answer pairs and beat BM25 on open-domain question answering \citep{Karpukhin2020DPR}. Khattab and Zaharia developed ColBERT, a late-interaction retriever that scores every query token against every document token for higher recall at a given rank \citep{Khattab2020ColBERT}; ColBERTv2 then compressed the representations and cut the index footprint by an order of magnitude \citep{Santhanam2022ColBERTv2}. Other encoders trade supervision for scale: Contriever transfers to zero-shot retrieval without supervised pairs \citep{Izacard2022Contriever}, SPLADE learns sparse lexical expansions over BERT logits \citep{Formal2021SPLADE}, ANCE mines hard negatives to improve dense retrievers \citep{Xiong2021ANCE}, and E5 and BGE scale contrastive pre-training into general-purpose embeddings that lead the MTEB leaderboard \citep{Wang2022E5, Xiao2024BGE, Guo2019NeuralIR}.

Hybrid retrieval combines sparse and dense signals. Ma and colleagues showed that weighted combinations of BM25 and DPR recover exact-match evidence that dense retrievers miss \citep{Ma2021CLEAR}. Cormack and colleagues introduced reciprocal rank fusion (RRF), a simple rank-based combiner that usually beats learned fusion when training signal is limited \citep{Cormack2009RRF}; later work confirmed that RRF is hard to beat without per-collection tuning \citep{Bruch2023Fusion}. Cross-encoder rerankers sit above the retriever: BERT as a listwise reranker \citep{Nogueira2019BERTRerank}, monoT5 for zero-shot transfer \citep{Nogueira2020MonoT5}, and LLM prompt-based rerankers that match supervised cross-encoders at higher cost \citep{Sun2023RankGPT, Xiong2025AFRRank, Rodrigues2025RiskSensitiveRanking}.

All of this work optimises relevance or recall within an unconstrained candidate space. None of it models admissibility. A ColBERTv2 retriever paired with an RRF combiner will happily return an A-Level physics question to an O-Level query if the topic matches. Educational retrieval has a stronger requirement: a document must be on-topic \emph{and} within the learner's curriculum scope. We build on hybrid retrieval and rank fusion, but we put a deterministic metadata gate in front of them, removing inadmissible documents before any ranking signal is computed.

\subsection{Retrieval-Augmented Generation}

Lewis and colleagues defined the retrieve-then-generate paradigm (RAG) that this paper sits inside \citep{Lewis2020RAG}. Several systems train retriever and generator together or change where retrieval happens: REALM trains them jointly \citep{Guu2020REALM}, Fusion-in-Decoder concatenates passages at the decoder \citep{Izacard2021FiD}, REPLUG adds a learned retriever at the prompt boundary of a black-box LLM \citep{Shi2024REPLUG}, and in-context RAG improves accuracy at inference time without fine-tuning \citep{Ram2023InContextRAG}.

Later RAG variants add reasoning or self-correction. IRCoT interleaves retrieval with chain-of-thought reasoning \citep{Trivedi2023IRCoT}; Self-RAG trains the model to decide when to retrieve and when to critique \citep{Asai2024SelfRAG}; RAPTOR retrieves over LLM-built hierarchical summaries \citep{Sarthi2024RAPTOR}; Corrective RAG rewrites or drops poorly grounded passages after retrieval \citep{Yan2024CRAG}; and Adaptive-RAG routes queries by predicted complexity \citep{Jeong2024AdaptiveRAG}. Surveys report that nearly all published variants improve retrieval \emph{quality} within an unconstrained pool rather than the \emph{admissibility} of what is retrieved \citep{Gao2023RAGSurvey, Fan2024RAGSurvey, Lin2025STrajRAG, Sun2025RetrieverGeneratorVerification, Yang2026LLMGuidedLegalIR}.

Table~\ref{tab:related_positioning} positions CHSR-RRF against five representative advanced RAG systems. Each improves retrieval, reasoning, or critique quality, and each assumes a document is eligible to be retrieved if it is semantically relevant. CHSR-RRF breaks that assumption: it treats curriculum scope as a hard pre-retrieval constraint and works only on the subset of the corpus that survives the gate. The method is orthogonal to, and composable with, any of the five systems in the table. A curriculum-gated Self-RAG or a curriculum-gated Corrective RAG is a strictly stronger system.

\subsection{Retrieval Benchmarks and Evaluation}

Standard retrieval benchmarks measure topical relevance, not scope. MS MARCO remains the default training corpus for neural retrievers \citep{Bajaj2016MSMARCO}; Natural Questions pairs real Google queries with Wikipedia answers \citep{Kwiatkowski2019NQ}; the TREC Deep Learning tracks built reusable test collections \citep{Craswell2020TRECDL}; BEIR measured zero-shot transfer across 18 tasks and showed dense retrievers do not uniformly beat BM25 out of domain \citep{Thakur2021BEIR}; MTEB broadened evaluation to 56 tasks \citep{Muennighoff2023MTEB}; and KILT unified knowledge-intensive NLP over a shared Wikipedia snapshot \citep{Petroni2021KILT, Aldayel2021StanceDetection, Azad2019QueryExpansion}.

Evaluation specific to RAG is younger. RAGAs scores faithfulness, answer relevance, and context precision without references \citep{Es2024RAGAS}; RGB probes four RAG failure modes \citep{Chen2024RGB}; CRAG is a 4{,}409-question benchmark with mock retrieval APIs \citep{Yang2024CRAG}; and industry guidance mirrors these efforts \citep{NVIDIA2025RAGEval}. None of these annotates queries or documents with hierarchical metadata that would reveal whether retrieved evidence is \emph{structurally} appropriate for the asker. BEIR measures topical precision, MTEB measures embedding geometry, RAGAs and RGB measure generator grounding, and CRAG measures noise robustness. The question ``did the system retrieve an A-Level physics passage when the student is in Form 4?'' cannot be answered with any of them.

A related concern, training-set contamination, has its own literature: mandatory per-benchmark contamination measurement \citep{Sainz2023Contamination}, strategies to stop test data leaking into training \citep{Jacovi2023StopUploading}, time-travel probes for memorised answers \citep{Golchin2024TimeTravel}, surveys of pervasive overlap \citep{Deng2024Contamination}, and atomic leakage-detection suites \citep{Xu2024Benbench}. Our notion of curriculum leakage is related but distinct: contamination measures training-set overlap, while curriculum leakage measures retrieval-time scope violations. Both reduce evaluation validity, and both are invisible to standard precision/recall metrics.

\subsection{Educational Information Retrieval and Tutoring}

Early intelligent tutoring systems used rule-based dialogue and knowledge tracing (Nwana, 1990; VanLehn, 2011), and surveys argued that AI-driven education needs structured curriculum models rather than shallow dialogue \citep{Luckin2016AIEd, ZawackiRichter2019AIEd, Baker2016StupidTutors}. The LLM era changed the surface but not the underlying problem: studies catalogue classroom opportunities and risks \citep{Kasneci2023ChatGPTEducation}, flag the near-total absence of retrieval-grounded tutoring evaluations \citep{Denny2024CSEdGenAI}, and document classroom patterns that require retrieval over course materials \citep{Mollick2023AssigningAI}. Industry tutors such as Khanmigo and Duolingo Max ship retrieval-grounded tutoring at scale, but their curriculum constraints are opaque to outside researchers \citep{KhanAcademy2023, Duolingo2023}. An educational RAG framework that couples retrieval with code interpreters for STEM QA still measures final-answer correctness rather than retrieval-time scope \citep{Lu2025EducationalQA}.

The gap this paper closes is the absence of an operational measure of retrieval-time curriculum fidelity. A tutor that looks helpful may be quietly pulling A-Level evidence into an O-Level answer, and the student's test score will not reveal why. Concurrent with our work, Rao proposed a Metadata-First Enterprise Architecture for governed AI systems, which enforces a governance gate before semantic search \citep{Rao2026MFEA}. MFEA and CHSR-RRF reach the same structural insight from opposite domains: MFEA frames governance at the enterprise-compliance level, CHSR-RRF at the curriculum level. MFEA is a framework proposal without empirical evaluation; CHSR-RRF is, to our knowledge, the first system to pair a formal curriculum-admissibility constraint with an ablation-verified \mbox{$4.6\times$} leakage reduction (\mbox{$p<0.001$}) and publicly released evaluation artifacts (Table~\ref{tab:related_positioning}).

\begin{table}[htbp]
\centering
\small
\caption{Positioning of CHSR-RRF relative to advanced RAG architectures.}
\label{tab:related_positioning}
\setlength{\tabcolsep}{3pt}
\resizebox{\linewidth}{!}{%
\begin{tabular}{@{}L{0.18\linewidth}L{0.24\linewidth}L{0.20\linewidth}L{0.24\linewidth}@{}}
\toprule
\textbf{System} & \textbf{Retrieval Strategy} & \textbf{Curriculum Awareness} & \textbf{Cost Model} \\
\midrule
IRCoT & Iterative retrieval + CoT & None & Multi-step LLM \\
Self-RAG & Learned retrieval + reflection & None & Fine-tuned model \\
RAPTOR & LLM-built hierarchy & Emergent (LLM cost) & Tree build + retrieval \\
CRAG & Post-hoc correction & None & Evaluator inference \\
Adaptive RAG & Complexity routing & None & Variable \\
\textbf{CHSR-RRF} & \textbf{Pre-gated hybrid + RRF} & \textbf{Hard metadata gate} & \textbf{Deterministic (no LLM)} \\
\bottomrule
\end{tabular}
}
\end{table}

CERB fills the gap the benchmarks above leave open. It annotates every query with a curriculum tuple, labels which evidence units are structurally admissible, and measures leakage as a first-class metric alongside nDCG and recall. The benchmark is not tied to Cameroon: it is a transferable protocol (hierarchy-aware relevance labels, leakage annotations, slice-level decomposition, and a failure taxonomy) that can be instantiated on any national examination system with structured metadata (UK GCSE and A-Level, Indian CBSE, French Baccalaur\'{e}at, West African WASSCE). The Cameroon GCE instantiation is the first concrete instance; cross-system instantiation is left as future work (Section~\ref{sec:conclusion}). To our knowledge, no prior work formulates retrieval as a constrained admissibility problem with explicit leakage tracking and auditable relaxation.

These prior studies improve retrieval quality but leave admissibility unaddressed. The next section describes the data, the method, and the benchmark we build to make admissibility measurable.

\section{Materials and Methods}
\label{sec:method}

This section presents the corpus, the CERB benchmark, the problem formulation, the CHSR-RRF retrieval pipeline, the ablation variants, the metrics, and the experimental setup, in that order. Following the reviewer's suggestion, the experimental setup is presented here rather than in a separate section, so the methodology reads as one continuous block.

\subsection{Corpus and Source Documents}
\label{sec:corpus_and_quality}

The evaluation corpus contains 74{,}018 evidence chunks indexed from over 2{,}000 source documents across 11 of the 14 GCE subjects, at both Ordinary and Advanced Levels. The source collection includes over ten years of past GCE examination papers, examiner and expert solutions, official syllabi from all three curriculum authorities, regional mock examinations (used as preparation for the national GCE exam), progression sheets, schemes of work, textbooks, workbooks, lesson notes, and lesson plans. Each subject carries a GCE Board subject code (e.g., 0580 for O-Level Physics, 0780 for A-Level Physics), which the retrieval gate uses for deterministic scope enforcement. Table~\ref{tab:corpus_profile} profiles the four largest subjects. Physics and Chemistry have the strongest metadata completeness and are the primary subjects in the CERB cases.

\begin{figure}[htbp]
\centering
\includegraphics[width=\linewidth]{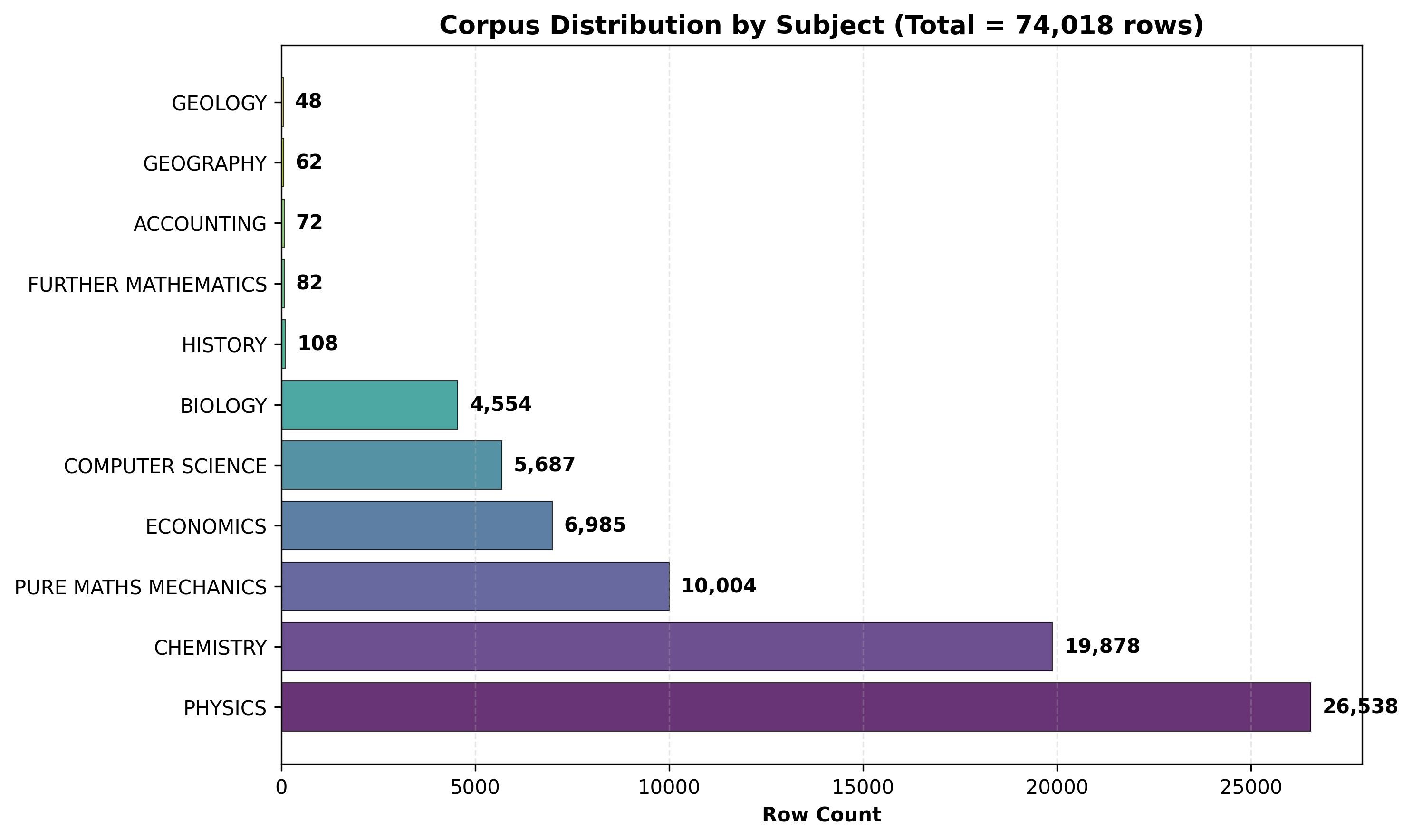}
\caption{Indexed corpus composition by subject ($n = 74{,}018$ chunks). Physics and Chemistry dominate (35.9\% and 26.8\%), reflecting the focus of the CERB cases on these two subjects. Five tail subjects (history, further mathematics, accounting, geography, geology) contribute 372 chunks combined (0.5\%), leaving several benchmark slices highly sensitive to missing-corpus coverage and metadata gaps.}
\label{fig:subject_distribution}
\end{figure}

\begin{figure}[htbp]
\centering
\includegraphics[width=\linewidth]{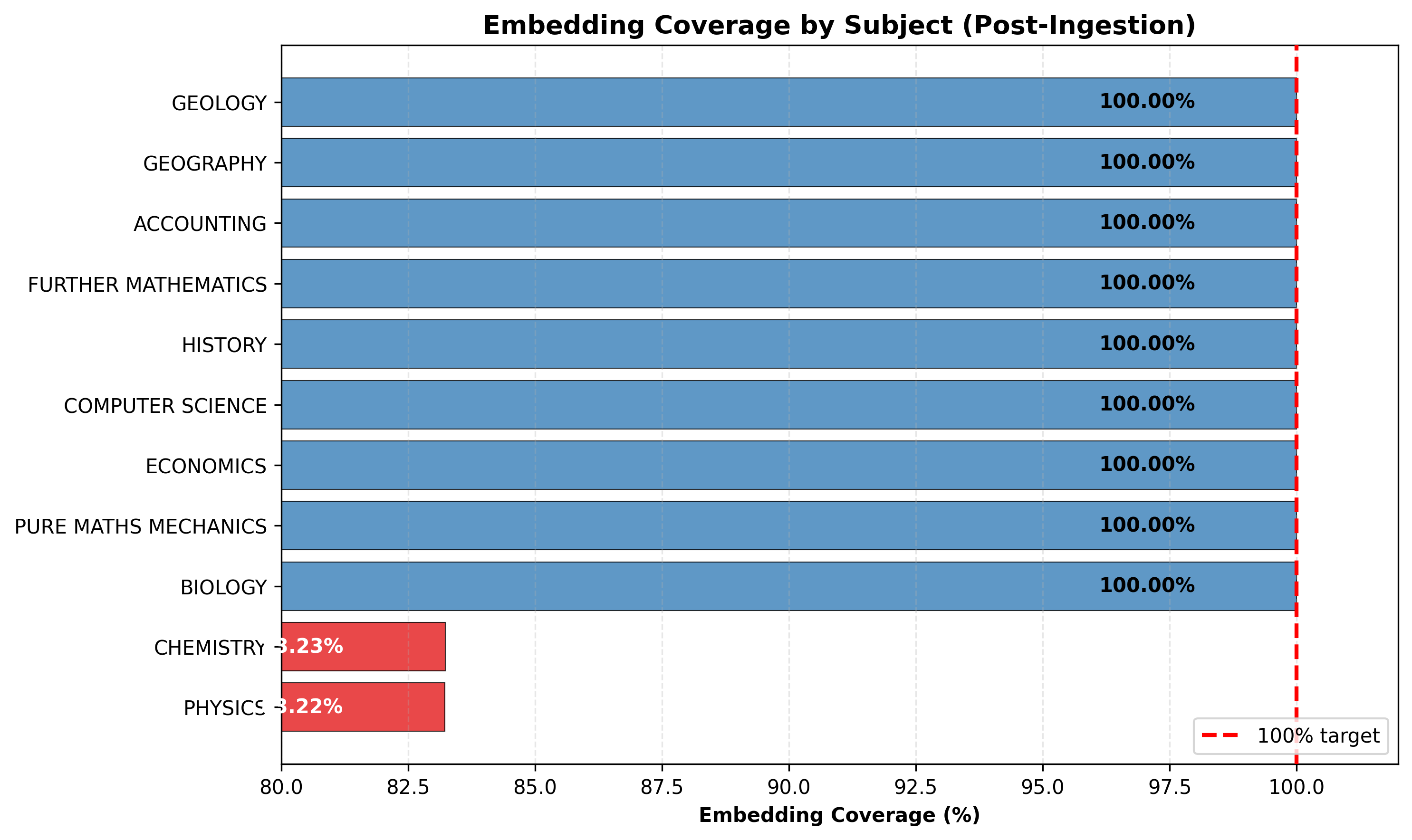}
\caption{Embedding coverage by subject after the V6 cleanup (target: 100\%, dashed line). Nine of eleven subjects reach the target; Physics (83.22\%) and Chemistry (83.23\%) carry residual NULL embeddings inherited from a Datalab extraction batch and addressed by the in-flight backfill.}
\label{fig:embedding_coverage}
\end{figure}

\begin{figure}[htbp]
\centering
\includegraphics[width=\linewidth]{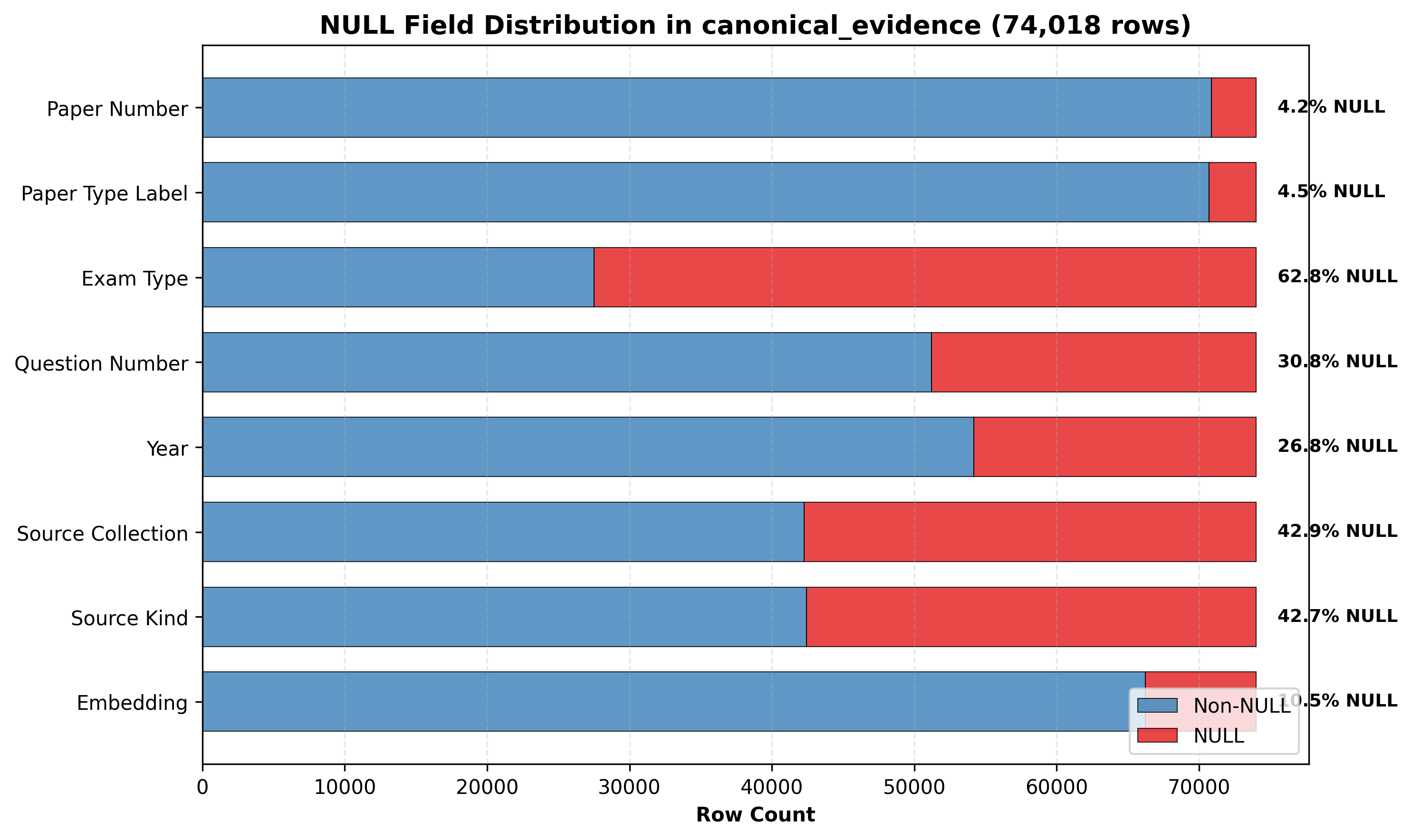}
\caption{NULL distribution across the eight metadata fields most relevant to curriculum gating ($n = 74{,}018$). Subject and level are 100\% populated (the gate is reliable on its primary keys), while \texttt{exam\_type} (62.85\% NULL), \texttt{source\_kind} (42.69\% NULL), \texttt{question\_number} (30.81\% NULL), and \texttt{year} (26.79\% NULL) carry residual gaps. Each NULL field directly bounds the achievable exact-scope success on the corresponding CERB cases and motivates the in-flight metadata backfill before the next ablation pass.}
\label{fig:null_distribution}
\end{figure}

\paragraph{Snapshot provenance.}
All corpus statistics in this paper are frozen against a canonical evaluation snapshot taken from the \code{canonical\_evidence} table after the V6 data-quality cleanup pass and archived in \code{code/results/corpus_snapshot_stats.md}. A later truncation-and-reingest cycle (driven by a metadata-first ingestion overhaul described in a companion paper) has since reduced the live database to a narrow KB~v2 slice. Table~\ref{tab:corpus_profile}, Fig.~\ref{fig:subject_distribution}, Fig.~\ref{fig:embedding_coverage}, and Fig.~\ref{fig:null_distribution} are therefore reproducible only from the frozen snapshot, not from the current live database. The frozen snapshot is the canonical evaluation corpus for this paper and the paired 61-case pilot ablation (Section~\ref{par:pilot_subset}).

\begin{table}[htbp]
\centering
\small
\caption{Corpus profile: evidence chunks per subject in the indexed evaluation corpus (74{,}018 total). Counts captured from a live query against the canonical evidence table after the V6 data-quality cleanup pass.}
\label{tab:corpus_profile}
\setlength{\tabcolsep}{4pt}
\begin{tabular}{@{}lr@{}}
\toprule
\textbf{Subject} & \textbf{Evidence Chunks} \\
\midrule
Physics & 26{,}538 \\
Chemistry & 19{,}878 \\
Pure Mathematics with Mechanics & 10{,}004 \\
Economics & 6{,}985 \\
Computer Science & 5{,}687 \\
Biology & 4{,}554 \\
Other (5 subjects) & 372 \\
\bottomrule
\end{tabular}
\end{table}

\paragraph{Data quality baseline.}\label{sec:data_quality}
The corpus underwent a documented V6 data-quality audit before the frozen evaluation run. Two facets matter for interpreting the ablation: embedding coverage (whether dense retrieval can return any candidate at all) and metadata completeness (whether the curriculum gate can match a chunk).

\paragraph{Embedding coverage.}
Of the 74{,}018 indexed chunks, 66{,}232 carry an OpenAI \texttt{text-embedding-3-small} vector (89.48\% coverage). The 7{,}786 (10.52\%) NULL-embedding rows are concentrated in two subjects: Physics (4{,}453 NULL, 83.22\% coverage) and Chemistry (3{,}333 NULL, 83.23\% coverage). The other nine subjects have 100\% embedding coverage. Fig.~\ref{fig:embedding_coverage} shows per-subject coverage and Fig.~\ref{fig:null_distribution} the NULL distribution across other metadata fields.

\subsection{The CERB Benchmark}

\paragraph{Educational context.}
CERB is grounded in the Cameroon General Certificate of Education (GCE) examination system. Cameroon runs two parallel national education systems: an English-language system following the British GCE model, and a French-language system following the French Baccalaur\'{e}at model. The GCE system, administered by the Cameroon GCE Board, covers the entire nation and examines students at two levels: Ordinary Level (O-Level) and Advanced Level (A-Level). It covers 14 subjects across sciences, arts, and applied disciplines, each examined at both levels (O-Level spans Forms 1--5; A-Level covers Lower and Upper Sixth). Each subject is examined through multiple papers: Paper 1 (multiple-choice), Paper 2 (structured and essay), and, in science subjects, Paper 3 (practical). Three curriculum authorities govern content at different stages: the Competency-Based Approach (CBA) in Forms 1 and 2, the National curriculum from the Ministry of Secondary Education, and the GCE Board examination syllabus. Official syllabus documents from all three authorities keep retrieval constraints aligned with the actual curriculum rather than with arbitrary metadata labels. This hierarchy (authority, subject, level, paper, year, question number) provides natural metadata for constrained retrieval, yet no existing system uses it as a formal retrieval gate.

\paragraph{Benchmark design.}
CERB is designed to evaluate retrieval in curriculum-constrained exam preparation. Each query--evidence pair carries:

\begin{enumerate}
    \item relevance labels
    \item curriculum leakage labels
    \item ambiguity markers
    \item task type labels for exact retrieval, scope relaxation, synthesis, and adversarial evaluation
\end{enumerate}

Each CERB case is a structured record with an identifier, a natural-language query, a declared curriculum scope (subject, level, paper), a slice label, expected evidence types, exact expected metadata fields, an optional expected question number, and annotation notes. The benchmark is distributed as a machine-readable specification that can be loaded, validated, and re-executed deterministically. The design shows why CERB is stricter than generic topical relevance: one case asks for projectile-motion evidence under Physics, O-Level, Paper 2 constraints with an expected question number, while another tests whether the same scope can be recovered under French phrasing. Separating semantic usefulness from scope correctness is what makes curriculum leakage measurable.

\begin{figure}[htbp]
\centering
\resizebox{\linewidth}{!}{%
\begin{tikzpicture}[
  node distance=7mm and 7mm,
  box/.style={rectangle, draw, rounded corners=1pt, minimum height=8mm,
              minimum width=24mm, align=center, font=\scriptsize},
  grp/.style={rectangle, draw, thick, rounded corners=2pt,
              minimum height=22mm, minimum width=28mm, align=center,
              font=\scriptsize\bfseries, fill=black!4},
  res/.style={rectangle, draw, fill=black!8, rounded corners=1pt,
              minimum height=7mm, minimum width=30mm, align=center, font=\scriptsize},
  arr/.style={-{Stealth[length=2mm]}, thick}
]
  \node[grp] (cases) {126 CERB cases\\7 slices};
  \node[box, right=of cases] (runner) {Ablation\\runner};
  \node[box, above right=4mm and 14mm of runner] (vA) {Variant A (ungated)};
  \node[box, right=of runner]                   (vC) {Variant C (gated)};
  \node[box, below right=4mm and 14mm of runner] (vE) {Variant E (+relax)};
  \node[box, right=20mm of vC] (retr) {Retrieval\\engine};
  \node[box, right=of retr] (score) {Scoring\\nDCG, Recall, Leakage, ESS};
  \node[res, below=of score] (perslice) {Per-slice results};
  \node[res, right=of perslice] (boot) {Bootstrap\\significance};

  \draw[arr] (cases)--(runner);
  \draw[arr] (runner)--(vA);
  \draw[arr] (runner)--(vC);
  \draw[arr] (runner)--(vE);
  \draw[arr] (vA.east)--(retr.west);
  \draw[arr] (vC)--(retr);
  \draw[arr] (vE.east)--(retr.west);
  \draw[arr] (retr)--(score);
  \draw[arr] (score)--(perslice);
  \draw[arr] (perslice)--(boot);
\end{tikzpicture}%
}
\caption{CERB evaluation pipeline. The 126 cases span seven slices: MCQ, structured, formula-heavy, cross-subject, bilingual, mark-scheme, and table-heavy. They are executed under five ablation variants (A: ungated, B: ungated with post-filter, C: gated hybrid, D: gated with reranker, E: full CHSR-RRF with relaxation). Every case emits a structured record scored on nDCG@$K$, Recall@$K$, leakage rate, and exact-scope success (ESS). Results are decomposed per slice and paired bootstrap significance tests are run on consecutive variant transitions.}
\label{fig:cerb_pipeline}
\end{figure}
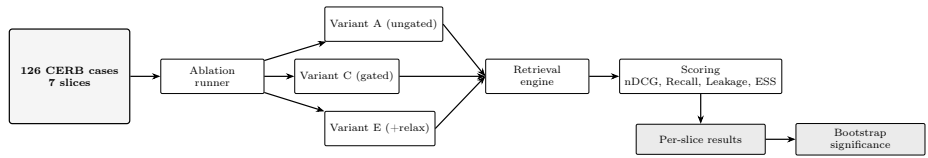

\paragraph{Annotation schema.}
Each CERB case is a structured record with the fields in Table~\ref{tab:cerb_schema}, designed to be machine-parseable and deterministically evaluable.

\begin{table}[htbp]
\centering
\small
\caption{CERB annotation schema per case.}
\label{tab:cerb_schema}
\setlength{\tabcolsep}{3pt}
\resizebox{\linewidth}{!}{%
\begin{tabular}{@{}L{0.22\linewidth}L{0.18\linewidth}L{0.48\linewidth}@{}}
\toprule
\textbf{Field} & \textbf{Type} & \textbf{Purpose} \\
\midrule
\code{id} & string & Unique case identifier (e.g., \code{mcq\_physics\_001}) \\
\code{query} & string & Natural-language query as a student would type it \\
\code{subject} & string & Ground-truth subject (e.g., \code{physics}, \code{chemistry}) \\
\code{level} & enum & \code{o\_level} or \code{a\_level} \\
\code{paper} & int $\mid$ null & Paper number (1, 2, 3) or null if unspecified \\
\code{slice} & enum & One of 7 task partitions (see Table~\ref{tab:cerb_composition}) \\
\code{expected\_chunk\_types} & list & Ingestion chunk types the query should retrieve \\
\code{expected\_fields} & dict & Exact metadata fields returned documents must match \\
\code{relevance\_keywords} & list & Terms indicating topical relevance \\
\code{expected\_question\_number} & string $\mid$ null & Specific question ID if applicable \\
\code{notes} & string & Debugging context and edge-case documentation \\
\bottomrule
\end{tabular}
}
\end{table}

\paragraph{Labeling policy.}
Cases were authored by domain experts familiar with Cameroon GCE exam structure and verified against the live corpus. The labeling rules are:
\begin{itemize}
    \item A retrieved document is \emph{relevant} if it addresses the topic in the query.
    \item A retrieved document is \emph{scope-correct} if its metadata matches all non-null fields in \code{expected\_fields}.
    \item A retrieved document is \emph{contaminated} if it is relevant to the topic but from the wrong question, paper, or year within the same subject and level.
    \item A retrieved document exhibits \emph{leakage} if it comes from a different subject or level than specified.
    \item A retrieval is an \emph{exact-scope success} only if at least one returned document matches both the expected metadata fields and, when specified, the expected question number.
\end{itemize}

\paragraph{Slices and composition.}
The 126 cases are partitioned into seven non-overlapping slices (Table~\ref{tab:cerb_composition}). \textbf{Each case is a single student query (one question), not a grouped set; the 126 cases are therefore 126 distinct queries, one per row of the slice counts in Table~\ref{tab:cerb_composition}.} CERB prioritises diagnostic depth over scale. Unlike large benchmarks that average over heterogeneous failure modes, CERB is built so that each case isolates a specific, interpretable failure condition (cross-level leakage, metadata absence, bilingual mismatch). This enables causal analysis of retrieval errors, which large but weakly annotated benchmarks cannot support. Slices are defined by query type and evidence format rather than by difficulty, so a failure on a given slice points to an identifiable system gap rather than to undifferentiated hard cases. The slice distribution is intentionally imbalanced: MCQ and structured questions dominate because they are the most common student query types, while bilingual, mark-scheme, and table-heavy cases serve as edge-case probes.

\begin{table}[htbp]
\centering
\caption{CERB benchmark composition: 126 cases across 7 curriculum slices. Each case is a single student query (one question); the counts below are therefore query counts.}
\label{tab:cerb_composition}
\begin{tabular}{@{}lr@{}}
\toprule
\textbf{Slice} & \textbf{Cases} \\
\midrule
Multiple-choice question (MCQ) retrieval & 35 \\
Structured questions & 20 \\
Formula-heavy queries & 15 \\
Cross-subject queries & 11 \\
Bilingual (EN/FR) queries & 15 \\
Mark-scheme retrieval & 15 \\
Table-heavy queries & 15 \\
\midrule
\textbf{Total} & \textbf{126} \\
\bottomrule
\end{tabular}
\end{table}

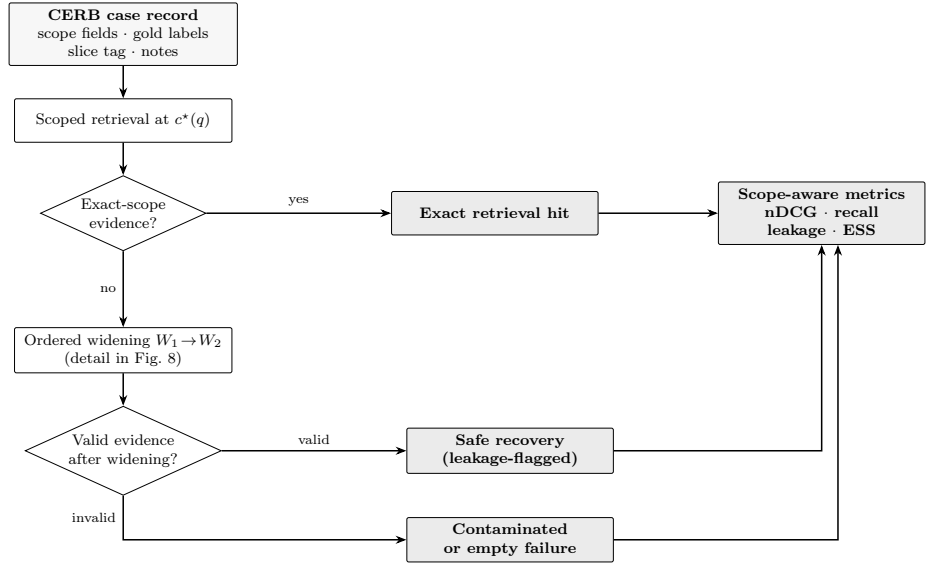
\begin{figure}[htbp]
\centering
\resizebox{\linewidth}{!}{%
\begin{tikzpicture}[
  node distance=6mm and 12mm,
  rec/.style={rectangle, draw, rounded corners=1pt, minimum height=8mm,
              minimum width=42mm, align=center, font=\footnotesize, fill=black!3},
  proc/.style={rectangle, draw, rounded corners=1pt, minimum height=8mm,
               minimum width=40mm, align=center, font=\footnotesize},
  dec/.style={diamond, draw, aspect=2.2, inner sep=0.5mm,
              align=center, font=\footnotesize},
  outbox/.style={rectangle, draw, fill=black!8, rounded corners=1pt,
              minimum height=8mm, minimum width=38mm, align=center, font=\footnotesize\bfseries},
  arr/.style={-{Stealth[length=2mm]}, thick},
  ylbl/.style={font=\scriptsize, midway, above, sloped},
  nlbl/.style={font=\scriptsize, midway, left}
]
  \node[rec] (rec) {\textbf{CERB case record}\\scope fields $\cdot$ gold labels\\slice tag $\cdot$ notes};
  \node[proc, below=of rec] (scoped) {Scoped retrieval at $c^\star(q)$};
  \node[dec, below=of scoped] (d1) {Exact-scope\\evidence?};
  \node[proc, below=14mm of d1] (widen) {Ordered widening $W_1\!\to\!W_2$\\(detail in Fig.~\ref{fig:scope_relaxation})};
  \node[dec, below=of widen] (d2) {Valid evidence\\after widening?};

  \node[outbox, right=34mm of d1] (hit) {Exact retrieval hit};
  \node[outbox, right=34mm of d2] (safe) {Safe recovery\\(leakage-flagged)};
  \node[outbox, below=8mm of safe] (bad) {Contaminated\\or empty failure};
  \node[outbox, right=22mm of hit] (metrics) {Scope-aware metrics\\nDCG $\cdot$ recall\\leakage $\cdot$ ESS};

  \draw[arr] (rec)--(scoped);
  \draw[arr] (scoped)--(d1);
  \draw[arr] (d1)--(hit) node[ylbl]{yes};
  \draw[arr] (d1)--(widen) node[nlbl]{no};
  \draw[arr] (widen)--(d2);
  \draw[arr] (d2)--(safe) node[ylbl]{valid};
  \draw[arr] (d2.south) |- (bad) node[pos=0.25, left, font=\scriptsize]{invalid};
  \draw[arr] (hit)--(metrics);
  \draw[arr] (safe.east) -| (metrics.south);
  \draw[arr] (bad.east) -| ([xshift=3mm]metrics.south);
\end{tikzpicture}%
}
\caption{CERB case structure and auditable scope-relaxation outcomes used in live evaluation. Each case record drives a scoped retrieval. If exact-scope evidence is found the trace ends in an exact hit; otherwise constraints are relaxed in the fixed, ordered widening sequence, yielding either a leakage-flagged safe recovery or a contaminated/empty failure. The diagram is drawn in vector form so the labels stay legible at print size.}
\label{fig:cerb_safe_widening}
\end{figure}

The benchmark is intended to separate three questions that are usually conflated: (1) did the retriever find relevant evidence? (2) did it stay within the valid curriculum scope? (3) if it widened scope, was that widening justified and controlled? Fig.~\ref{fig:cerb_safe_widening} shows how those decisions are made visible in the benchmark record and the widening trace. Widening is not hidden inside retrieval heuristics: each relaxation step is explicit, ordered, and auditable at case level.

\paragraph{Two-phase evaluation: pilot ablation and final-corpus measurements.}
\label{par:pilot_subset}
The empirical evidence is reported in two complementary phases against the same corpus snapshot. \textbf{Phase~1 (pilot ablation, \S\ref{sec:results}):} the six-variant paired-bootstrap ablation that establishes the gating-versus-filtering asymmetry is computed on a 61-case pilot subset of CERB, spanning all seven slices (MCQ~20, structured~10, formula-heavy~10, cross-subject~6, bilingual~5, mark-scheme~5, table-heavy~5). The pilot is sized for paired-bootstrap power on the headline gating effect ($A\!\to\!B$, $4.6\times$ leakage reduction, $p<0.001$), where the effect size dominates plausible sampling variance at $n=61$. \textbf{Phase~2 (final-corpus measurements, \S\ref{sec:postfix_validation}):} a single-variant \code{fts_only_scoped} measurement and a paired within-corpus probe are run against the full 126-case benchmark on the current corpus, isolating corpus-side normalisation effects from algorithmic effects. A six-variant rerun on the full 126 cases is deferred to a companion arXiv update pending corpus re-ingestion (\S\ref{sec:conclusion}).

\subsection{Problem Formulation and Metadata Hierarchy}

We formulate educational retrieval as a constrained admissibility problem. Given a query $q$ and a corpus $D$ of evidence units, each document $d \in D$ carries a curriculum tuple
\[
c(d) = (\text{subject}, \text{level}, \text{year}, \text{session}, \text{paper}, \text{content\_type}).
\]
The query induces a constraint tuple $c^\star(q)$, which may be fully or partially specified. The retrieval objective is to maximise relevance within the admissible subset:
\[
\max_{R \subseteq D_g(q)} \; \text{Rel}(R, q),
\qquad
D_g(q) = \{ d \in D \mid c(d) \models c^\star(q) \}.
\]

Two competing objectives follow. Relevance measures whether retrieved documents are semantically useful and is reported through nDCG@$K$ and Recall@$K$. Admissibility measures whether every retrieved document lies inside the valid curriculum scope, and is reported through leakage rate and exact-scope success. Standard RAG systems optimise relevance without enforcing admissibility, which is the source of the curriculum leakage described in Section~\ref{sec:intro}.

The contribution of CHSR-RRF is not any single component. Hybrid retrieval, reciprocal rank fusion, and cross-encoder reranking are all well established. The contribution is treating retrieval as a constrained admissibility problem, where the candidate pool is restricted \emph{before} ranking and relaxation is explicitly modelled, ordered, and audited (Section~\ref{sec:safe_widening}). This differs from existing systems that treat constraints, where they appear at all, as post-hoc filters on an unconstrained pool. The framing connects educational retrieval to constrained search (Baeza-Yates and Ribeiro-Neto, 2011) and domain-specific retrieval (Metzler et al., 2021) \citep{BaezaYates2011IR, Metzler2021Rethinking}, extending Broder's query taxonomy \citep{Broder2002Taxonomy} with a category whose correctness depends on structural validity in a domain hierarchy rather than on topical match alone. Table~\ref{tab:notation} (at the front of the paper) lists the symbols used throughout.

\subsection{The CHSR-RRF Retrieval Pipeline}

\begin{figure}[htbp]
\centering
\resizebox{\linewidth}{!}{%
\begin{tikzpicture}[
  node distance=7mm and 6mm,
  box/.style={rectangle, draw, rounded corners=1pt, minimum height=8mm,
              minimum width=18mm, align=center, font=\footnotesize},
  gate/.style={rectangle, draw, thick, fill=black!6, rounded corners=1pt,
               minimum height=8mm, minimum width=22mm, align=center, font=\footnotesize\bfseries},
  arr/.style={-{Stealth[length=2mm]}, thick},
  lbl/.style={font=\scriptsize\itshape, gray}
]
  \node[box] (q1)  {Query};
  \node[box, right=of q1] (e1)  {Embed};
  \node[box, right=of e1] (v1)  {Vector\\search};
  \node[box, right=of v1] (k1)  {Top-$k$};
  \node[box, right=of k1] (l1)  {LLM};
  \node[box, right=of l1] (a1)  {Answer};
  \draw[arr] (q1)--(e1); \draw[arr] (e1)--(v1); \draw[arr] (v1)--(k1);
  \draw[arr] (k1)--(l1); \draw[arr] (l1)--(a1);
  \node[lbl, above=0.5mm of v1] (bl) {Baseline RAG: relevance only};

  \node[box, below=11mm of q1] (q2)  {Query};
  \node[box, right=of q2] (h2)  {Extract\\scope};
  \node[gate, right=of h2] (g2)  {Curriculum\\gate};
  \node[box, right=of g2] (f2)  {BM25 $\oplus$\\Dense / RRF};
  \node[box, right=of f2] (r2)  {Rerank};
  \node[box, right=of r2] (w2)  {Relax\\(if $<k_{\min}$)};
  \node[box, right=of w2] (l2)  {LLM};
  \node[box, right=of l2] (a2)  {Answer};
  \draw[arr] (q2)--(h2); \draw[arr] (h2)--(g2); \draw[arr] (g2)--(f2);
  \draw[arr] (f2)--(r2); \draw[arr] (r2)--(w2); \draw[arr] (w2)--(l2); \draw[arr] (l2)--(a2);
  \node[lbl, below=1mm of g2] {CHSR-RRF: relevance $+$ admissibility};
\end{tikzpicture}%
}
\caption{Baseline RAG (top) retrieves by semantic relevance alone and lets the LLM handle any scope mismatch after the fact. CHSR-RRF (bottom) extracts the curriculum scope from the query, applies a hard metadata gate before any ranking signal is computed, then runs hybrid sparse-dense retrieval with reciprocal rank fusion, a deterministic reranker, and progressive scope relaxation only if the gated pool falls below $k_{\min}$.}
\label{fig:baseline_vs_chsrrrf}
\end{figure}
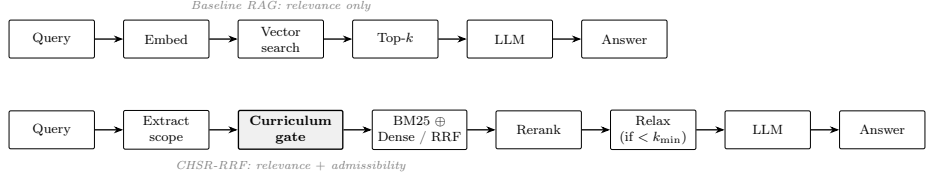

Fig.~\ref{fig:baseline_vs_chsrrrf} contrasts a baseline RAG pipeline with CHSR-RRF. The proposed method comprises five stages, each formally defined below.

\paragraph{Stage 1: Hard Curriculum Gate.}
The gate is a binary function that excludes documents whose metadata does not match the query's extracted curriculum scope:
\[
G(d, q) = \begin{cases}
1 & \text{if } s_d = s(q) \;\wedge\; \ell_d = \ell(q) \;\wedge\; (p(q) = \bot \;\vee\; p_d = p(q)) \\
0 & \text{otherwise}
\end{cases}
\]
where $\bot$ denotes an unspecified field. The gate requires exact match on subject and level, and exact match on paper number when specified. The gated candidate pool is $D_g(q) = \{d \in D \mid G(d,q) = 1\}$. This gate is implemented as SQL \code{WHERE} clauses on indexed metadata columns, executed at the database level before any embedding computation.

\paragraph{Stage 2: Dual-Path Retrieval.}
Within the gated pool $D_g(q)$, two retrieval paths execute in parallel:
\begin{align}
R_{\text{bm25}} &= \text{TopK}_{\text{bm25}}(q, D_g, k_1) \\
R_{\text{dense}} &= \text{TopK}_{\text{dense}}(\mathbf{e}(q), D_g, k_2)
\end{align}
BM25 (Best Match 25) uses PostgreSQL \code{tsvector} full-text search with OR-mode matching ($k_1 = 20$). Dense retrieval uses pgvector HNSW (Hierarchical Navigable Small World) cosine similarity ($k_2 = k$). Deterministic query expansion adds curriculum terms and synonyms, producing up to two additional candidate lists $R_{\text{bm25}}^{(e)}$ and $R_{\text{dense}}^{(e)}$.

\paragraph{Stage 3: Reciprocal Rank Fusion.}
All non-empty candidate lists are merged using reciprocal rank fusion (RRF) \citep{Cormack2009RRF}:
\[
\text{RRF}(d) = \sum_{r \in \mathcal{R}} \frac{1}{\kappa + \text{rank}_r(d)}
\]
where $\mathcal{R}$ is the set of candidate lists containing $d$, and $\kappa = 60$. We use $\kappa = 60$ because it is the standard default proposed by \citet{Cormack2009RRF}; RRF is known to be robust to this constant, so the choice does not materially affect the ranking.

\paragraph{Stage 4: Deterministic Reranker.}
A metadata-aware reranker adjusts the fused score:
\[
\text{score}(d, q) = \alpha \cdot \text{RRF}(d) + \beta \cdot \text{Align}(d, h) + \gamma \cdot \text{Rel}(d) - \delta \cdot \text{Red}(d)
\]
where Align$(d, h)$ is curriculum alignment (boosts for exact question number, correct content type, and correct document family), Rel$(d)$ is source reliability (official exam papers $>$ curated textbooks $>$ informal notes), and Red$(d)$ is a redundancy penalty from content-based deduplication (similarity threshold 0.75). No LLM inference is required. The expression above is the idealised additive form used for exposition; the production implementation (\texttt{backend\_api/src/features/retrieval/ranker.py}) realises the same policy as a multiplicative composition of 13 per-feature boost factors (range 1.05--1.40) and 3 penalty factors (0.55--0.75) applied to the RRF score. The two formulations are monotonically equivalent in ranking under a log transformation, so the ablation results are unaffected by the choice of aggregation; the multiplicative form is simply more convenient to tune per-factor.

\paragraph{Stage 5: Scope Relaxation.}
Defined in Section~\ref{sec:safe_widening} below.

\begin{figure}[htbp]
\centering
\includegraphics[width=\linewidth,height=0.78\textheight,keepaspectratio]{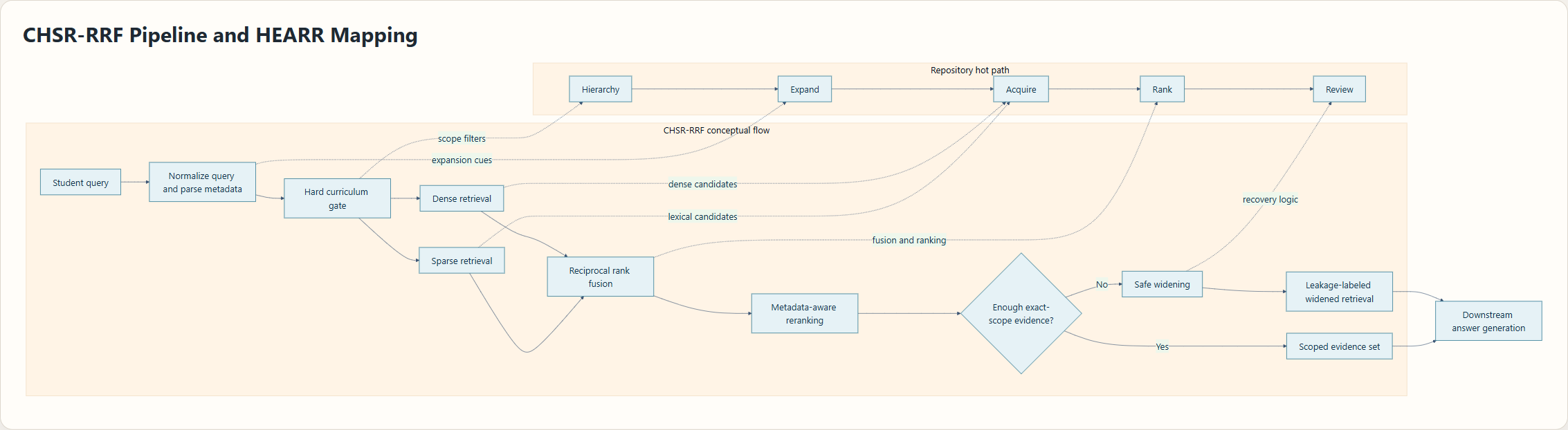}
\caption{CHSR-RRF retrieval pipeline and mapping to the HEARR-style implementation. Hard curriculum gating restricts the candidate set before sparse retrieval, dense retrieval, reciprocal rank fusion, and metadata-aware reranking. Progressive scope relaxation is attempted only after the scoped path fails to recover enough exact-scope evidence. The lower mapping shows the correspondence between the formal stages and the production runtime.}
\label{fig:chsr_rrf_method}
\end{figure}

CHSR-RRF is the formal retrieval policy; the production system instantiates it as a software runtime whose stage names map directly onto Stages~1--5 above.\footnote{The production runtime acronym is HEARR (Hierarchical, Expansion, Agentic dual-path, Rank fusion, Review-and-widen). The mapping is one-to-one: H~$\to$~Stage~1 (hierarchy gate), E~$\to$~query expansion in Stage~2, A~$\to$~dual-path retrieval in Stage~2, R~$\to$~Stages~3--4, R~$\to$~Stage~5. We use the formal CHSR-RRF naming throughout this paper; HEARR appears only in the released code package.} Fig.~\ref{fig:chsr_rrf_method} summarises the formal pipeline, and Algorithm~1 gives the procedure.

\paragraph{Algorithm.}

\begin{figure}[htbp]
\centering
\small
\fbox{\parbox{0.92\linewidth}{%
\textbf{Algorithm 1:} CHSR-RRF Retrieval \\[4pt]
\textbf{Input:} query $q$, corpus $D$, top-$k$ parameter $k$, min-results $k_{\min}$ \\
\textbf{Output:} ranked list $R$ of size $\leq k$ \\[4pt]
1: $h \leftarrow \text{extract\_hierarchy}(q)$ \hfill \textit{// Query understanding} \\
2: $D_g \leftarrow \{d \in D \mid G(d, q) = 1\}$ \hfill \textit{// Hard curriculum gate} \\
3: $q_e \leftarrow \text{expand}(q, h)$ \hfill \textit{// Curriculum-aware expansion} \\
4: $R_{\text{bm25}} \leftarrow \text{TopK}_{\text{bm25}}(q, D_g, k_1)$ \\
5: $R_{\text{dense}} \leftarrow \text{TopK}_{\text{dense}}(\mathbf{e}(q), D_g, k)$ \\
6: $R_{\text{bm25}}^{(e)} \leftarrow \text{TopK}_{\text{bm25}}(q_e, D_g, k_1)$ \hfill \textit{// If expansion produced terms} \\
7: $R_{\text{dense}}^{(e)} \leftarrow \text{TopK}_{\text{dense}}(\mathbf{e}(q_e), D_g, k)$ \\
8: $R \leftarrow \text{RRF}(R_{\text{bm25}}, R_{\text{dense}}, R_{\text{bm25}}^{(e)}, R_{\text{dense}}^{(e)};\; \kappa{=}60)$ \\
9: $R \leftarrow \text{rerank}(R, h)$ \hfill \textit{// Metadata boost + dedup} \\
10: $i \leftarrow 0$ \\
11: \textbf{while} $|R| < k_{\min}$ \textbf{and} $i < 2$ \textbf{do} \\
12: \quad $i \leftarrow i + 1$ \\
13: \quad $D_g \leftarrow \text{relax}(D_g, i)$ \hfill \textit{// Progressive relaxation} \\
14: \quad $R' \leftarrow \text{retrieve\_and\_rank}(q, q_e, D_g)$ \\
15: \quad $R \leftarrow \text{merge}(R, R')$;\; \text{label\_leakage}$(R, h)$ \\
16: \textbf{end while} \\
17: \textbf{return} $\text{TopK}(R, k)$
}}
\end{figure}

\subsubsection{Progressive Scope Relaxation}
\label{sec:safe_widening}

\paragraph{Formal leakage definition.}
A retrieved document $d$ exhibits \emph{leakage} with respect to query $q$ when it violates the original scope constraint:
\[
\text{leakage}(d, q) = \begin{cases}
\text{cross-topic} & \text{if } t_d \neq t(q) \;\wedge\; \ell_d = \ell(q) \\
\text{cross-level} & \text{if } \ell_d \neq \ell(q) \\
\text{cross-subject} & \text{if } s_d \neq s(q) \\
\bot & \text{otherwise (no leakage)}
\end{cases}
\]

Scope relaxation treats insufficient recall as a retrieval-state problem rather than a free licence to expand. The trigger is not a simple count threshold but a four-case decision tree that separates corpus sparsity from a genuine topic miss:

\begin{enumerate}
    \item \textbf{Zero results}: always widen. The scope constraint is too tight for the available corpus.
    \item \textbf{Results match the requested subject}: \textbf{skip widening}. The corpus is simply small for this scope; widening would add leakage without improving relevance.
    \item \textbf{Results do not match the requested subject}: widen. The dense retriever returned cross-subject content despite the scope filter, indicating a genuine topic miss.
    \item \textbf{No subject filter active}: widen on count alone, to preserve broad exploration for unscoped queries.
\end{enumerate}

Case~2 is the key insight: when $|\{d \in R(q) : s_d = s(q)\}| > 0$ but $|R(q)| < k_{\min}$, widening would replace on-topic results with off-topic results from a larger candidate pool. The decision tree short-circuits this counterproductive expansion.

When widening is triggered, constraints are relaxed in a fixed, auditable order:

\begin{enumerate}
    \item $W_1$: drop topic and subtopic constraints while keeping subject, level, authority, paper, and year. Leakage type introduced: \emph{cross-topic}.
    \item $W_2$: drop level and cycle constraints while keeping subject and authority. Leakage type introduced: \emph{cross-level}. This step is \textbf{blocked} when strict-scope keys are present (paper number, year, question number, document type, content type, or mark-scheme flag), preventing cross-level leakage on precise queries.
    \item $W_3$ (operational): if filtered retrieval is still empty, a final unfiltered lexical fallback is permitted as an explicit escape hatch. This fallback is \textbf{not} counted as benchmark success.
\end{enumerate}

At each widening step, the full retrieval pipeline (dual-path $\to$ RRF $\to$ rerank) is re-executed against the relaxed candidate pool. Maximum widening attempts: 2. Each widened result carries an explicit leakage label so quality and scope violation can be analysed jointly.

\begin{figure}[htbp]
\centering
\begin{tikzpicture}[
  node distance=6mm and 10mm,
  box/.style={rectangle, draw, rounded corners=1pt, minimum height=7mm,
              minimum width=34mm, align=center, font=\scriptsize},
  dec/.style={diamond, draw, aspect=2, inner sep=0.4mm,
              align=center, font=\scriptsize},
  term/.style={rectangle, draw, fill=black!6, rounded corners=1pt,
               minimum height=7mm, minimum width=28mm, align=center, font=\scriptsize\bfseries},
  arr/.style={-{Stealth[length=2mm]}, thick},
  yeslbl/.style={font=\scriptsize, above, sloped},
  nolbl/.style={font=\scriptsize, below, sloped}
]
  \node[box] (q) {Run gated pipeline at scope $c^\star(q)$};
  \node[dec, below=of q] (d1) {$|R| \geq k_{\min}$?};
  \node[term, right=15mm of d1] (ok) {Return results};
  \node[dec, below=of d1] (d2) {Results match\\requested subject?};
  \node[term, right=15mm of d2] (skip) {Skip widening,\\return small result};
  \node[box, below=of d2] (w1) {$W_1$: drop topic $+$ subtopic};
  \node[dec, below=of w1] (d3) {$|R| \geq k_{\min}$?};
  \node[term, right=15mm of d3] (ok2) {Return (flag: cross-topic)};
  \node[box, below=of d3] (w2) {$W_2$: drop level (blocked if strict keys)};
  \node[term, below=of w2] (out) {Return (flag: cross-level or empty)};

  \draw[arr] (q)--(d1);
  \draw[arr] (d1.east)--(ok) node[yeslbl, midway] {yes};
  \draw[arr] (d1)--(d2) node[nolbl, midway, right=1mm] {no};
  \draw[arr] (d2.east)--(skip) node[yeslbl, midway] {yes};
  \draw[arr] (d2)--(w1) node[nolbl, midway, right=1mm] {no};
  \draw[arr] (w1)--(d3);
  \draw[arr] (d3.east)--(ok2) node[yeslbl, midway] {yes};
  \draw[arr] (d3)--(w2) node[nolbl, midway, right=1mm] {no};
  \draw[arr] (w2)--(out);
\end{tikzpicture}%
\caption{Progressive scope-relaxation flowchart. Widening is triggered only when the gated pool is below $k_{\min}$ and the small pool does not already match the requested subject. The $W_2$ step is blocked when the query carries strict-scope keys (paper, year, question number, document type, mark-scheme flag), preventing cross-level leakage on precise queries. Every relaxed result is emitted with an explicit leakage label so downstream audits can separate admissible recall from relaxation-induced leakage.}
\label{fig:scope_relaxation}
\end{figure}
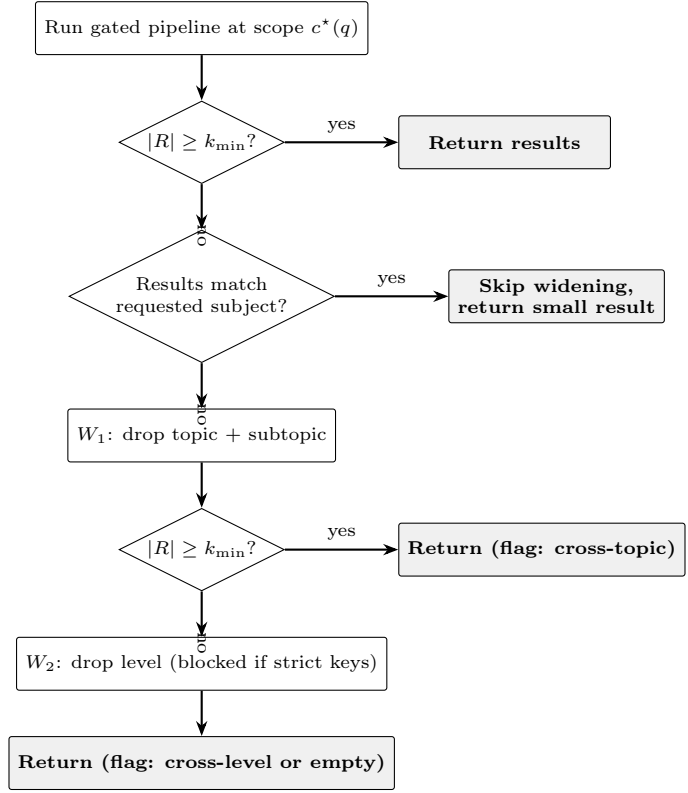

\subsection{Ablation Variants}

The ablation compares six retrieval variants, each adding one capability to isolate its marginal contribution:

\begin{enumerate}
    \item[\textbf{A.}] \textbf{Dense-only (scoped)}: cosine similarity over 1536-dimensional embeddings with a hard curriculum metadata gate. No lexical retrieval, no fusion.
    \item[\textbf{B.}] \textbf{Hybrid ungated}: BM25 lexical retrieval + dense retrieval + reciprocal rank fusion, but \emph{no} metadata gate. This is the unconstrained baseline.
    \item[\textbf{B+.}] \textbf{Hybrid post-filtered}: same as B, with curriculum metadata filters applied \emph{after} retrieval. This diagnostic baseline tests whether post-hoc filtering can substitute for pre-retrieval gating.
    \item[\textbf{C.}] \textbf{Hybrid gated}: same as B, with the hard curriculum gate applied before retrieval. This isolates the gating effect.
    \item[\textbf{D.}] \textbf{Gated + reranked}: C plus the deterministic metadata-aware reranker and content-based deduplication (threshold 0.75).
    \item[\textbf{E.}] \textbf{Full CHSR-RRF}: D plus curriculum-aware query expansion and progressive scope relaxation with leakage tracking.
\end{enumerate}

All variants retrieve the top $K = 8$ documents per query against the same indexed corpus (74{,}018 evidence chunks across 11 subjects, with Physics and Chemistry comprising the majority of indexed content). Retrieval is executed against a PostgreSQL database with pgvector HNSW indexing and BM25 full-text search. All runs completed with zero errors.

\paragraph{Mapping to common baseline designs.}
The six variants give empirical form to policy alternatives that prior RAG work has discussed conceptually. \textbf{B} (\emph{hybrid ungated}) stands in for an ``ungated retrieval + downstream prompt-level constraint'' design: the retriever surfaces any topically relevant document, and any scope enforcement happens elsewhere. \textbf{B+} (\emph{hybrid post-filtered}) stands in for a ``post-filter + retry'' design that ranks by relevance and then applies metadata filtering, mirroring the corrective-RAG \citep{Yan2024CRAG} family at the metadata layer rather than the relevance layer. \textbf{D} (\emph{gated + reranked}) stands in for ``post-rerank with curriculum signals,'' where the deterministic reranker plays the role of an LLM-based critic (as in Self-RAG, \citealp{Asai2024SelfRAG}) at zero LLM cost. We do not run the LLM-based systems themselves because each makes architectural assumptions (LLM critic, multi-hop retrieval planning, hierarchical summarisation) that are orthogonal to whether pre-retrieval curriculum gating helps; running their full pipelines would not isolate the gate effect. The empirical baselines we do run (B, B+, D) isolate exactly the design dimensions those systems vary along.

\subsection{Metrics}

We report six metrics that separate retrieval quality from curriculum fidelity:

\begin{itemize}
    \item \textbf{Retrieval quality}: Precision@K, Recall@K, and normalized discounted cumulative gain (nDCG@K).
    \item \textbf{Curriculum fidelity}: leakage rate (proportion of results violating curriculum scope), contamination rate (correct scope but wrong question), and exact-scope success (ESS: at least one result matching all expected metadata fields and question number).
    \item \textbf{Efficiency}: end-to-end latency per query, including network round-trips.
\end{itemize}

\subsection{Experimental Setup and Statistical Method}

We report paired bootstrap confidence intervals ($N = 1000$, seed = 42) for consecutive ablation pairs. For exact-scope success, which is binary and sparse, we report $p$-values with the caution that the 61-case pilot provides limited power for detecting small differences between intermediate variants. All ablation results are derived from a single frozen evaluation run over the 61-case CERB pilot subset at $K = 8$ against the canonical corpus snapshot. The 126-case CERB specification, the 61-case pilot case list, the ablation runner, and per-case result artifacts are all versioned in a companion code package. Re-executing the ablation against the same corpus and embedding model produces identical results. The CERB dataset and evaluation code will be released upon publication.

\paragraph{Annotator-consistency check (method).}\label{sec:iaa}
CERB's 126 cases were authored by a single annotator. True inter-annotator agreement (IAA) requires two independent human raters and is deferred to future work. As a conservative substitute we ran an \emph{LLM-judge consistency check}: a deterministic rule-based judge re-derives each case's \texttt{subject}, \texttt{level}, and \texttt{paper\_number} from the \emph{query text and notes alone}, without seeing \texttt{expected\_fields}, \texttt{expected\_question\_number}, or any other annotator-assigned metadata. The judge emits a label only when the query contains an explicit signal (e.g., ``O-Level'', ``Paper 1'', ``MCQ'', or a subject keyword); otherwise it abstains. Cohen's~$\kappa$ is computed against the annotator labels on the full 126 cases (artefact: \code{llm_judge_consistency.json}). The results of this check are reported with the other findings in Section~\ref{sec:iaa_results}.

With the data, the method, and the evaluation set up, the next section presents the results and discusses what they mean.

\section{Results and Discussion}
\label{sec:results}

\subsection{Main Ablation Results}

\paragraph{Headline result.}
Applying the same curriculum metadata constraints \emph{after} retrieval (Variant~B+) scores \textbf{zero} on every retrieval metric (R@8, nDCG@8, ESS all 0), while applying them \emph{before} retrieval (Variant~C) recovers \mbox{nDCG@8 = 0.5529} and cuts leakage by $4.6\times$ relative to the ungated hybrid baseline~B ($0.7596 \to 0.1653$, $p < 0.001$). The two configurations differ only in enforcement point; the asymmetry is the paper's central claim. Table~\ref{tab:full_live_cerb} reports the full six-variant ablation behind this contrast, and Table~\ref{tab:slice_recall} breaks it down by slice.

\begin{table}[htbp]
\centering
\small
\caption{Ablation results on the 61-case pilot subset of CERB ($n = 61$ cases spanning all 7 slices, $K = 8$). ESS = exact-scope success. Bold indicates best value per column.}
\label{tab:full_live_cerb}
\setlength{\tabcolsep}{4pt}
\resizebox{\linewidth}{!}{%
\begin{tabular}{@{}L{0.21\linewidth}rrrrrrr@{}}
\toprule
\textbf{Variant} & \textbf{P@8} & \textbf{R@8} & \textbf{nDCG@8} & \textbf{Leakage} & \textbf{Contam.} & \textbf{ESS} & \textbf{Latency (ms)} \\
\midrule
A: Dense-only scoped & 0.0143 & 0.0410 & 0.4874 & 0.0969 & 0.0764 & 0.0492 & 911 \\
B: Hybrid ungated & 0.0020 & 0.0164 & 0.2577 & 0.7596 & 0.2602 & 0.0164 & 15760 \\
B+: Post-hoc filtered & 0.0000 & 0.0000 & 0.0000 & 0.0000 & 0.0000 & 0.0000 & 17799 \\
C: Hybrid gated & 0.0139 & 0.0410 & 0.5529 & 0.1653 & 0.1284 & 0.0492 & 2361 \\
D: Gated + reranked & 0.0139 & 0.0410 & 0.5495 & 0.1646 & 0.1294 & 0.0492 & 1754 \\
E: Full CHSR-RRF & 0.0214 & 0.0492 & 0.5508 & 0.2593 & 0.1288 & 0.0656 & 2645 \\
\bottomrule
\end{tabular}
}
\end{table}

\begin{table}[htbp]
\centering
\small
\caption{Per-slice Recall@8 across the five primary ablation variants (B+ omitted; all zeros). Three slices (bilingual, mark-scheme, table-heavy) record zero recall across all variants. Cross-subject is the only slice where the full CHSR-RRF pipeline (E) uniquely recovers recall.}
\label{tab:slice_recall}
\setlength{\tabcolsep}{4pt}
\begin{tabular}{@{}lrrrrr@{}}
\toprule
\textbf{Slice} & \textbf{A} & \textbf{B} & \textbf{C} & \textbf{D} & \textbf{E} \\
\midrule
MCQ & 0.050 & 0.050 & 0.050 & 0.050 & 0.050 \\
Structured & 0.100 & 0.000 & 0.100 & 0.100 & 0.100 \\
Formula-heavy & 0.050 & 0.000 & 0.050 & 0.050 & 0.050 \\
Cross-subject & 0.000 & 0.000 & 0.000 & 0.000 & 0.083 \\
Bilingual & 0.000 & 0.000 & 0.000 & 0.000 & 0.000 \\
Mark-scheme & 0.000 & 0.000 & 0.000 & 0.000 & 0.000 \\
Table-heavy & 0.000 & 0.000 & 0.000 & 0.000 & 0.000 \\
\bottomrule
\end{tabular}
\end{table}

\begin{figure}[htbp]
\centering
\includegraphics[width=\linewidth]{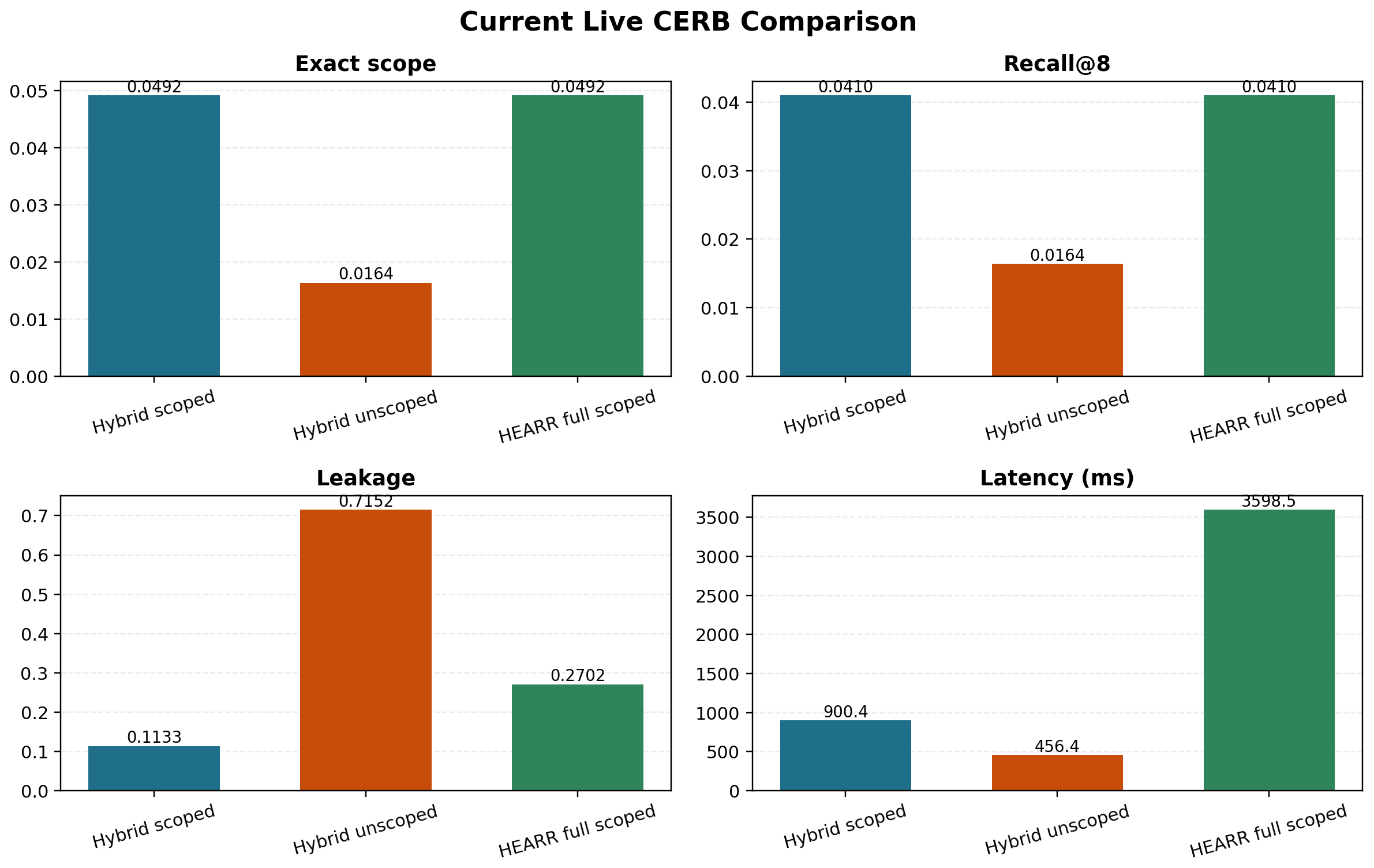}
\caption{Six-variant ablation on the 61-case CERB pilot ($n = 61$, $K = 8$). Curriculum gating (A$\to$C) reduces leakage while maintaining recall; post-hoc filtering (B+) collapses to zero; scope relaxation (D$\to$E) recovers cross-subject recall at the cost of higher leakage and latency.}
\label{fig:live_cerb_comparison}
\end{figure}

\begin{figure}[htbp]
\centering
\includegraphics[width=\linewidth]{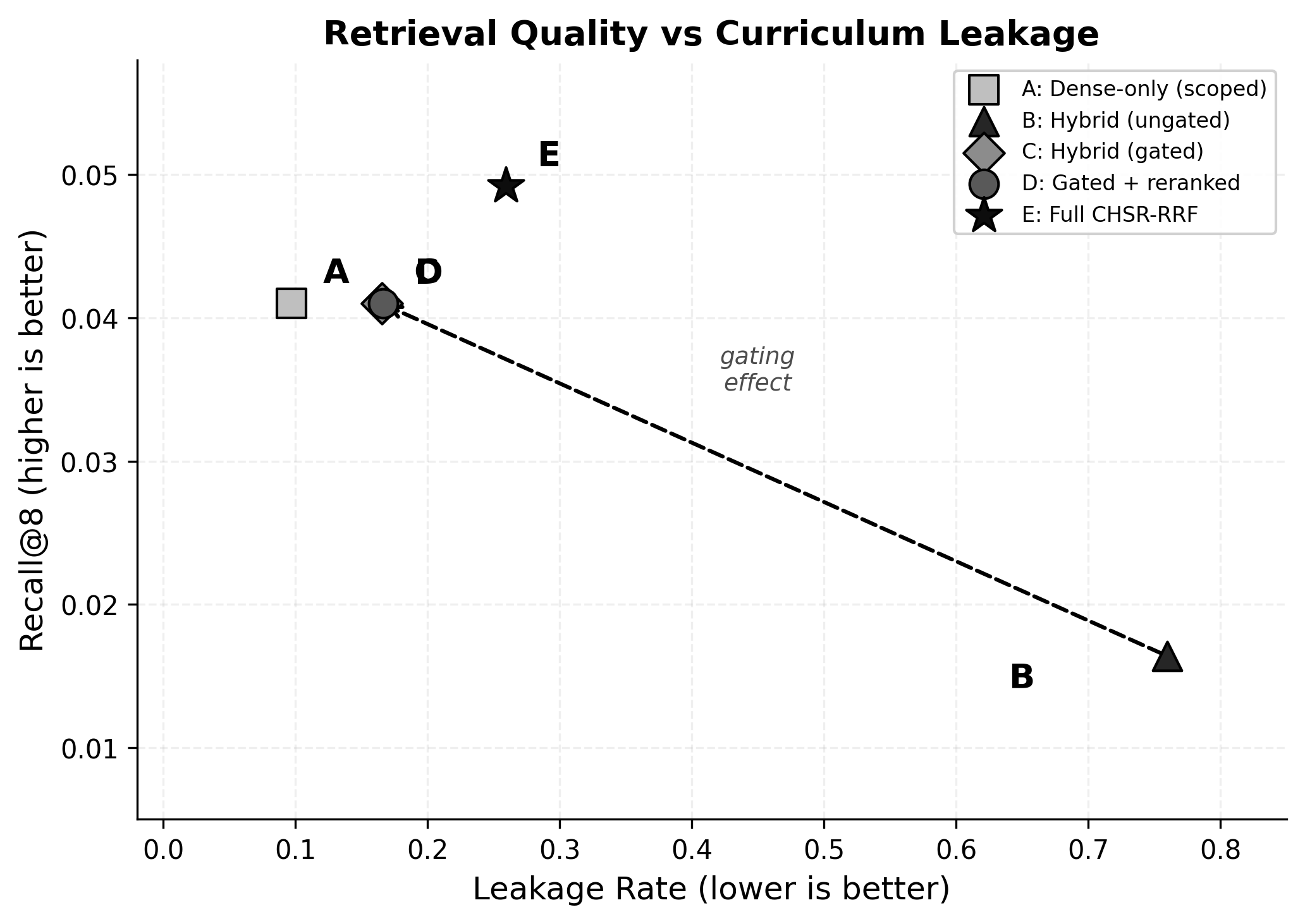}
\caption{Recall@8 versus leakage rate for all six ablation variants. The dashed arrow shows the gating effect: adding curriculum constraints (B$\to$C) sharply reduces leakage while improving recall. Variant E trades higher leakage for the only non-zero cross-subject recovery.}
\label{fig:leakage_recall}
\end{figure}

\begin{figure}[htbp]
\centering
\includegraphics[width=\linewidth]{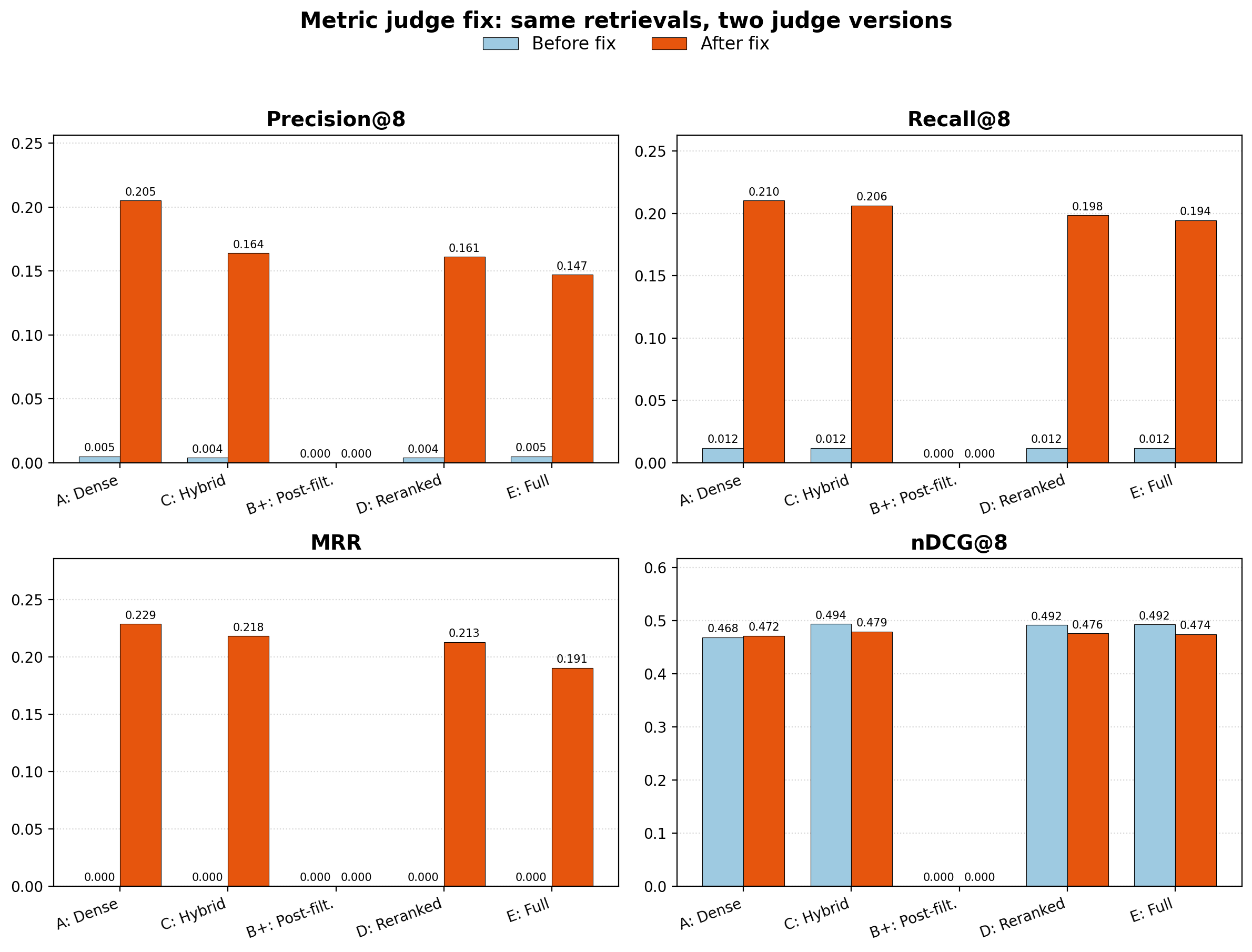}
\caption{Offline rescoring of the same 630 case results under the broken and the V6-fixed metric judge, shown as a 2$\times$2 panel of Precision@8, Recall@8, MRR, and nDCG@8. Identical retrievals, two judge implementations, side by side. The fix to \texttt{judge\_exact\_scope} and the \texttt{recall\_at\_k} fallback (which credits exact-scope documents the chunker had mislabeled) produces a 29--41$\times$ jump in Precision@8 and a 16--17$\times$ jump in Recall@8 across the non-postfiltered variants, with MRR moving from a literal 0.000 to 0.19--0.23. The numbers reported throughout this section are from the fixed judge.}
\label{fig:metric_fix}
\end{figure}

The four findings below decompose the contrast previewed above.

\paragraph{Finding 1: Pre-retrieval gating is not replaceable by post-hoc filtering.}
Pre-retrieval gating changes candidate-pool composition in a way post-hoc filtering cannot recover. Variant~B+ tests the alternative directly: retrieve from the full corpus (Variant~B) and discard wrong-scope documents afterwards. The result is total collapse: B+ scores zero on every metric (Recall@8 = 0, nDCG@8 = 0, ESS = 0). When 76\% of B's ungated results violate curriculum scope, post-filtering removes nearly all documents, leaving empty or near-empty result sets. Pre-gating (Variant~C) fills all $K$ slots from the admissible pool, achieving nDCG@8 of \mbox{0.5529} and ESS of \mbox{0.0492}. The B+ $\to$ C difference is statistically significant for both recall ($p = 0.039$) and leakage ($p < 0.001$). The asymmetry is structural: post-hoc correction cannot recover documents that were never retrieved. This is the load-bearing empirical claim of the paper.

\paragraph{Finding 2: Gating accounts for the largest single effect in the pipeline.}
Removing the curriculum gate (A$\to$B) increases leakage from \mbox{0.0969} to \mbox{0.7596} and drops nDCG@8 from \mbox{0.4874} to \mbox{0.2577}, the largest single transition in the ablation ($p < 0.001$ on bootstrap test, Table~\ref{tab:bootstrap_significance}). Together with Finding~1, this isolates pre-retrieval gating as the only mechanism in the ablation that materially controls leakage at acceptable recall.

\paragraph{Finding 3: Hybrid fusion improves ranked quality within scope.}
Adding full-text search (FTS) and RRF to the gated dense baseline (A$\to$C) raises nDCG@8 from \mbox{0.4874} to \mbox{0.5529} while keeping the same exact-scope success (\mbox{0.0492}). Leakage rises moderately from \mbox{0.0969} to \mbox{0.1653}. The deterministic reranker layered on top of fusion (C$\to$D) does not improve retrieval quality at this sample size: $p = 1.000$ on the paired bootstrap.

\paragraph{Finding 4: Scope relaxation uniquely recovers cross-subject recall, at a cost.}
The full CHSR-RRF pipeline (E) is the only variant that achieves non-zero cross-subject recall (\mbox{0.083} vs \mbox{0.000} for all other variants, Table~\ref{tab:slice_recall}), raising exact-scope success from \mbox{0.0492} to \mbox{0.0656} (a 33\% improvement on the pilot). The recovery comes with a measurable trade-off: leakage rises from \mbox{0.1646} to \mbox{0.2593} and latency from \mbox{1754}\,ms to \mbox{2645}\,ms. Three slices (bilingual, mark-scheme, table-heavy) remain at zero recall across all five variants; we treat this in Section~\ref{sec:root_causes} as a diagnostic finding about the corpus, not a method failure. Fig.~\ref{fig:live_cerb_comparison} visualises the trade-off.

\begin{table}[htbp]
\centering
\small
\caption{Bootstrap significance tests ($N=1000$, seed=42) for consecutive ablation pairs. $p$-values below 0.05 indicate statistically significant differences.}
\label{tab:bootstrap_significance}
\setlength{\tabcolsep}{4pt}
\resizebox{\linewidth}{!}{%
\begin{tabular}{@{}L{0.22\linewidth}rrr@{}}
\toprule
\textbf{Pair} & \textbf{Recall@K $p$} & \textbf{Exact scope $p$} & \textbf{Leakage $p$} \\
\midrule
A $\to$ B & 0.000 & 0.000 & 0.000 \\
B $\to$ B+ & 0.000 & 0.000 & 0.000 \\
B+ $\to$ C & 0.039 & 0.039 & 0.000 \\
C $\to$ D & 1.000 & 1.000 & 0.336 \\
D $\to$ E & 0.342 & 0.342 & 0.000 \\
\bottomrule
\end{tabular}
}
\end{table}

\subsection{Uncertainty and Slice-Level Analysis}
\label{sec:analytical_results}

\paragraph{Paired uncertainty.}
Table~\ref{tab:bootstrap_significance} sharpens the interpretation. The A$\to$B pair (removing the gate) is significant across all three metrics ($p < 0.001$), confirming that curriculum gating is the dominant effect. The B$\to$B+ pair (post-hoc filtering of ungated results) is also significant ($p < 0.001$): post-filtering produces significantly worse results than even the ungated baseline because it removes most documents without recovering admissible alternatives. The B+$\to$C pair (pre-gating versus post-filtering) is significant for both recall ($p = 0.039$) and leakage ($p < 0.001$), giving direct statistical evidence that pre-gating and post-filtering are not equivalent strategies. The C$\to$D pair (adding the reranker) shows no significant effect ($p \geq 0.336$). The D$\to$E pair (scope relaxation) produces a significant leakage increase ($p < 0.001$) but the recall improvement does not reach significance ($p = 0.342$).

\paragraph{Slice-level coverage and failure modes.}
The per-slice breakdown (Table~\ref{tab:slice_recall}) exposes structured failure modes that aggregate metrics obscure. Three slices (bilingual, mark-scheme, table-heavy) record zero recall across all five variants. Cross-subject is the only slice where CHSR-RRF (E) uniquely recovers recall (0.083), while all other variants remain at zero. The ungated variant (B) loses structured and formula-heavy recall that the gated variants retain, confirming that unconstrained retrieval degrades not only scope fidelity but also recall on specific content types. The primary bottleneck is corpus coverage and metadata completeness, not retrieval algorithm design.

\paragraph{Cross-subject analysis.}
The cross-subject slice (11 cases) is the only partition where the full CHSR-RRF pipeline (E) uniquely recovers recall. Variant E achieves Recall@8 of \mbox{0.083} on this slice, while all other variants, including the gated hybrid (C) and gated + reranked (D), remain at zero. This confirms that scope relaxation, which loosens the hard gate when the scoped pipeline produces insufficient results, is the mechanism responsible for cross-subject recovery. However, the recovery comes with increased leakage (\mbox{0.2593} for E vs \mbox{0.1646} for D) and substantially higher latency (\mbox{2645}\,ms vs \mbox{1754}\,ms). The right reading is not that cross-subject retrieval is solved, but that the benchmark exposes a slice where controlled relaxation helps while still leaving a clear error surface.

\subsection{Annotator-Consistency Findings}
\label{sec:iaa_results}

\begin{figure}[htbp]
\centering
\includegraphics[width=\linewidth]{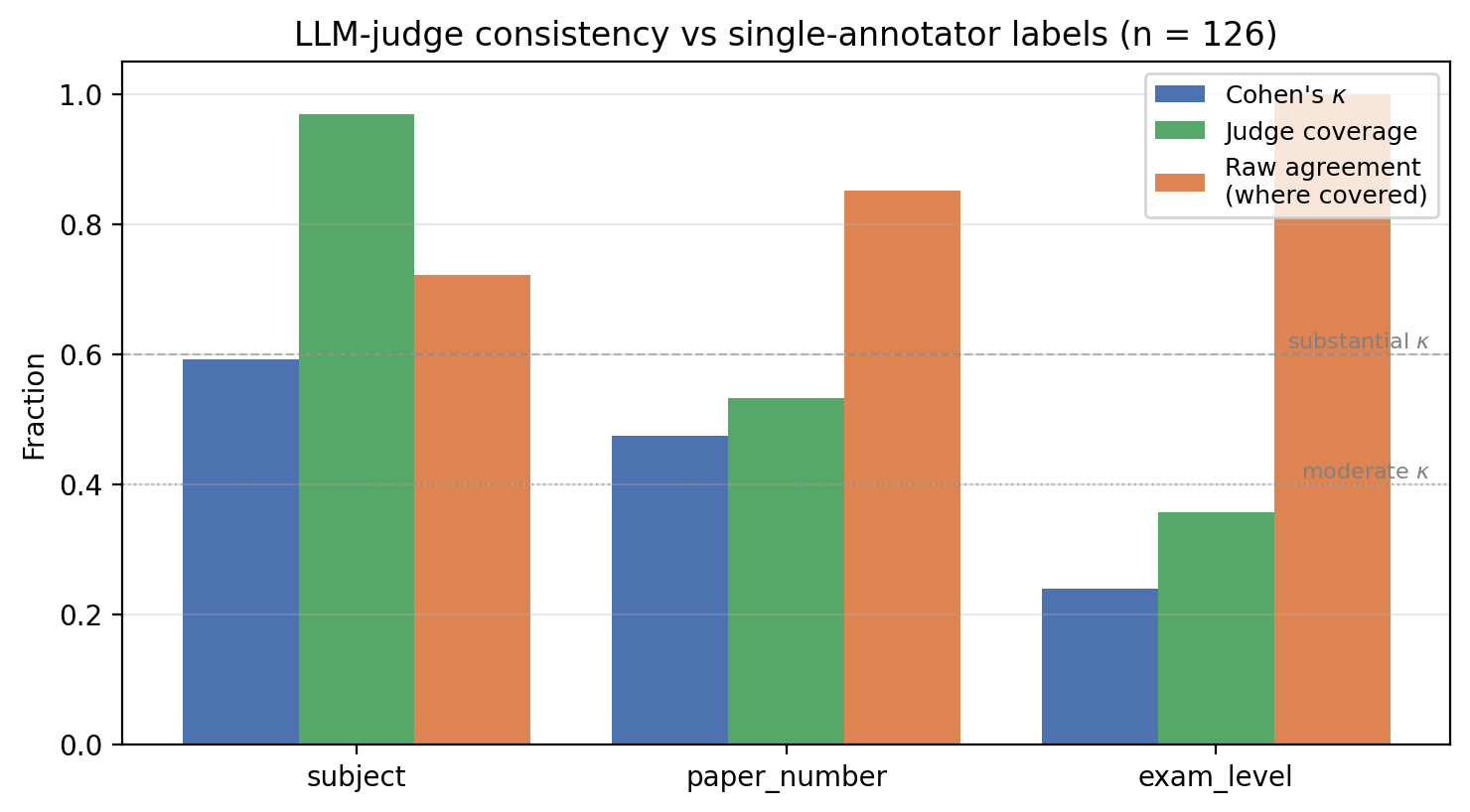}
\caption{LLM-judge consistency check on the 126-case CERB benchmark. Cohen's $\kappa$, judge coverage (fraction of cases where the rule-based judge produces a label rather than abstaining), and raw agreement on the covered subset. Subject is moderately well-determined from query text; \texttt{exam\_level} is rarely made explicit in queries, so the annotator's level label reflects external context rather than query content.}
\label{fig:iaa_kappa}
\end{figure}

We now report the LLM-judge consistency check whose method is described in Section~\ref{sec:iaa}. Table~\ref{tab:iaa_consistency} gives the agreement figures and Fig.~\ref{fig:iaa_kappa} visualises them.

\begin{table}[htbp]
\centering
\small
\caption{LLM-judge consistency against the single-annotator CERB labels ($n = 126$). Coverage = fraction of cases the judge labelled (abstention fraction: $1 - \text{coverage}$). Raw agreement is measured over the covered subset. Subject is well-determined by the query; level is rarely explicit in queries and mostly supplied from annotator context.}
\label{tab:iaa_consistency}
\setlength{\tabcolsep}{6pt}
\begin{tabular}{@{}lrrr@{}}
\toprule
\textbf{Field} & \textbf{Cohen's~$\kappa$} & \textbf{Coverage} & \textbf{Raw agreement} \\
\midrule
\texttt{subject}         & 0.592 & 0.968 & 0.721 (88/122) \\
\texttt{paper\_number}   & 0.474 & 0.532 & 0.851 (57/67) \\
\texttt{exam\_level}     & 0.238 & 0.357 & 1.000 (45/45) \\
\bottomrule
\end{tabular}
\end{table}

Three observations follow. First, subject is moderately well-identified from query text alone ($\kappa = 0.59$, 72\% raw agreement over 122 covered cases); disagreements concentrate in the bilingual and cross-subject slices, where the query deliberately spans or obscures subject boundaries. Second, \texttt{paper\_number} is identifiable when the query contains ``Paper 1/2'' or ``MCQ/structured'' cues (53\% coverage, 85\% agreement where covered). Third, and most importantly, \texttt{exam\_level} is rarely made explicit in query text (35\% coverage); the annotator supplied the level label from knowledge of the source corpus rather than from query content. This is a legitimate gap in the current CERB protocol: queries that implicitly carry curriculum-level context are under-constrained as retrieval test inputs when the level label is exactly what the retriever must match. The $\kappa$ values should therefore not be read as evidence of annotator unreliability; they are evidence that the current query-writing convention leaves level and paper under-specified, which future CERB revisions can address by requiring each query to encode the curriculum scope it intends to test. A two-human-rater IAA study on an expanded CERB is planned for a follow-up paper.

\subsection{Root-Cause Analysis of Zero-Recall Slices}
\label{sec:root_causes}

\begin{figure}[htbp]
\centering
\includegraphics[width=\linewidth]{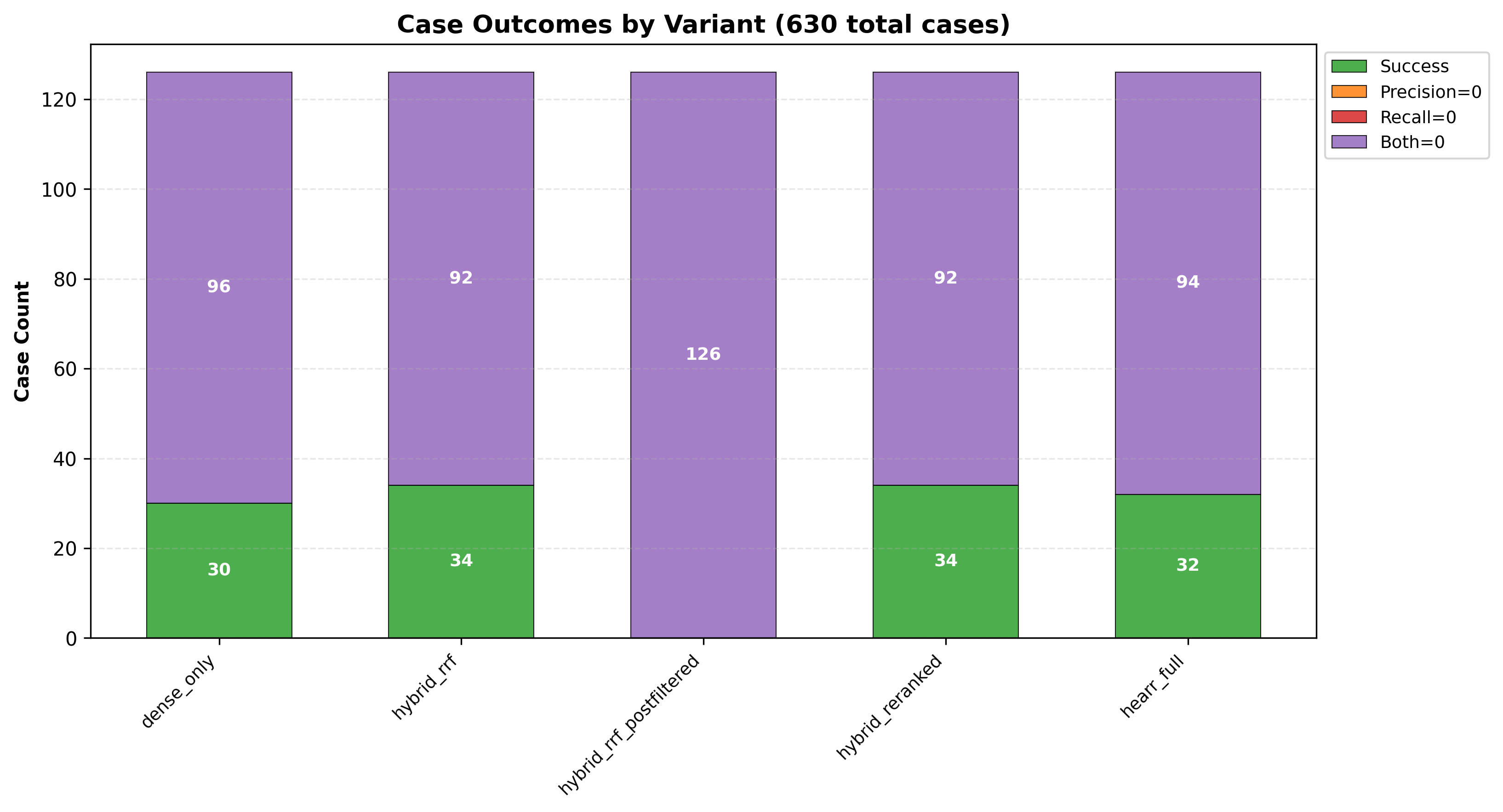}
\caption{Per-variant failure-mode breakdown across the 61-case CERB pilot under the V6-fixed judge. ``Both fail'' indicates non-empty retrievals where both Precision@8 and Recall@8 score zero, which isolates genuine retrieval hardness from empty candidate sets. The B+ post-filtered variant is omitted because it returns empty result sets on all 61 pilot cases.}
\label{fig:failure_modes}
\end{figure}

Absolute performance is low: the best variant achieves \mbox{0.0656} exact-scope success and three of seven slices record zero recall across all variants. CERB is designed to expose exactly this boundary. It strips away the cushion that aggregate metrics provide and forces every retrieval failure to be classified by cause. Fig.~\ref{fig:failure_modes} decomposes per-variant outcomes into four exclusive categories (success, precision-only failure, recall-only failure, both-fail), making the residual hardness explicit at the case level. We classify each zero-recall slice by root cause below.

We classify each failing case into three categories: \emph{data gap} (the expected evidence does not exist in the corpus), \emph{query gap} (the query preprocessing cannot handle the input form), or \emph{method gap} (a design limitation of CHSR-RRF itself). Table~\ref{tab:root_causes} summarises the diagnosis; the post-fix recall numbers in Table~\ref{tab:postfix_delta} validate it slice by slice (preprocessing-classified slices recover; data-gap-classified slices stay at zero).

\begin{table}[htbp]
\centering
\small
\caption{Root-cause classification of the five lowest-performing CERB slices (36 cases). Data gaps dominate: 21 cases fail because the target subject has near-zero ingested content, not because of method design.}
\label{tab:root_causes}
\setlength{\tabcolsep}{3pt}
\resizebox{\linewidth}{!}{%
\begin{tabular}{@{}L{0.16\linewidth}rL{0.30\linewidth}L{0.14\linewidth}L{0.24\linewidth}@{}}
\toprule
\textbf{Slice} & \textbf{Cases} & \textbf{Primary Root Cause} & \textbf{Classification} & \textbf{Fix Path} \\
\midrule
Bilingual & 5 & Biology and economics not ingested; French queries produce zero FTS matches & Data + Query gap & Ingest missing subjects; bilingual \code{simple} FTS index shipped post-eval (migration~0032); FR\,$\rightarrow$\,EN expansion future work \\
Mark-scheme & 5 & \code{document\_type} not labelled during ingestion; biology not ingested & Data gap (metadata) & Ingest missing subjects; canonical \code{document\_type} vocabulary shipped post-eval (commit~\code{09746337}) \\
Table-heavy & 5 & 3/5 target subjects with no data; no \code{data\_table} chunk types & Data gap & Ingest missing subjects; add table metadata \\
Cross-subject & 11 & 7/11 target subjects sparse; single-subject gate blocks multi-subject queries & Data + Method gap & Ingest subjects; multi-subject intent detection \\
Formula-heavy & 10 & Formula tokens split by BM25; no formula normalisation; 2/10 target math (sparse data) & Query + Data gap & Corpus-side \code{normalize_formula_text} shipped post-eval (commit~\code{09746337}); query-side normalisation future work \\
\bottomrule
\end{tabular}
}
\end{table}

\paragraph{Bilingual slice.}
All five bilingual cases target Biology or Economics, subjects with near-zero ingested content. The hard curriculum gate correctly filters to \code{subject=BIOLOGY} or \code{subject=ECONOMICS}, producing empty candidate sets regardless of query quality. Independently, at evaluation time the two French-language queries received zero lexical matches because the FTS index used \code{to_tsvector('english', ...)} stemming that dropped French tokens, and no bilingual query expansion was applied. A post-evaluation migration (\code{0032_has_figures_and_bilingual_fts.sql}) adds a parallel language-agnostic \code{simple} FTS index that addresses the English-stemming side of the gap; FR\,$\rightarrow$\,EN query expansion remains future work. Dense embeddings (text-embedding-3-small) provide some cross-lingual capability, but without French content in the corpus, even semantic similarity cannot bridge the gap.

\paragraph{Mark-scheme slice.}
The five mark-scheme cases expect documents with \code{document\_type=mark\_scheme}. At evaluation time, the ingestion pipeline processed mark-scheme PDFs but labelled their content as regular solution or answer chunks, leaving the \code{document\_type} field effectively unpopulated, so mark-scheme-specific filters produced empty candidate sets even when the underlying content existed in the corpus. A post-evaluation ingestion fix (commit \code{09746337}) adds a canonical document-type vocabulary and normaliser (\code{normalize_document_type} in \code{evidence_normalizers.py}) that collapses the \code{mark\_scheme}, \code{marking\_scheme}, and \code{marking\_guide} aliases to a canonical \code{pamphlet\_solutions} label. Re-evaluation on the expanded corpus is deferred to a companion arXiv update.

\paragraph{Table-heavy slice.}
Three of five table-heavy cases target Biology, Economics, or Computer Science, all subjects with near-zero data. The remaining two cases (Physics, Chemistry) expose a secondary gap: the ingestion pipeline does not create dedicated \code{data\_table} chunk types, and table content embedded within broader chunks is poorly represented in both BM25 (table cells form incoherent text) and dense retrieval (embedding models handle tabular data poorly).

\paragraph{Cross-subject slice.}
The cross-subject slice is the only one that exposes a genuine method limitation. The query understanding module extracts a single subject, and the hard gate filters to that subject. For queries like ``GCE past paper multiple subjects physics chemistry biology,'' relevant documents from non-primary subjects are excluded by design. This affects about 5 of 11 cases and represents a deliberate trade-off: the hard gate prevents leakage but blocks legitimate cross-subject exploration. Future work could address this with a query-intent classifier that bypasses the hard gate for discovery queries.

\paragraph{Formula-heavy slice.}
At evaluation time, formula queries (e.g., ``KE = 1/2 mv\textsuperscript{2}'') were handled by both BM25 (operators split tokens) and dense retrieval (formula text has weak semantic similarity to LaTeX-rendered content) without any normalisation. A post-evaluation ingestion fix (commit \code{09746337}) introduces corpus-side \code{normalize_formula_text} (in \code{evidence_text_utils.py}) that strips LaTeX math delimiters, normalises Unicode super- and subscripts to explicit \texttt{\^{}N}/\texttt{\_N} tokens, and drops \texttt{\textbackslash vec\{\}} wrappers from the search-text used by BM25, mitigating the operator-splitting problem on the corpus side. Query-side formula normalisation remains future work.

\paragraph{Summary.}
The dominant failure mode is data gap, not method gap: 21 of 36 cases in the five lowest-performing slices fail because the target subject has near-zero ingested content. A further 10 cases fail due to preprocessing gaps (no bilingual expansion, no formula normalisation). Only 5 cases expose a genuine CHSR-RRF design limitation (single-subject gating blocking legitimate cross-subject queries). This distinction is critical for directing improvement effort: over 85\% of current failures are addressable through corpus expansion and preprocessing without algorithmic redesign. Illustrative failure cases are detailed in Appendix~\ref{sec:failure_appendix}.

\subsection{Post-Fix Lower-Bound Measurement on the Full 126-Case Benchmark}
\label{sec:postfix_validation}

The root-cause analysis in Table~\ref{tab:root_causes} predicts that formula-heavy, bilingual, and cross-subject failures are preprocessing-addressable (query/data gaps) while mark-scheme and table-heavy failures are corpus-coverage gaps. We shipped two corpus-side fixes: (i)~\code{normalize_formula_text} in \code{evidence_text_utils.py}, which strips LaTeX math delimiters and normalises Unicode super/subscripts before BM25 indexing, and (ii)~migration \code{0032_has_figures_and_bilingual_fts.sql}, which adds a parallel language-agnostic \code{simple} FTS index. With both in place we ran an \code{fts_only_scoped} variant (BM25-only + hard curriculum gate, dense path deliberately ablated) against the full 126-case CERB benchmark. This variant is not one of the six primary ablation configurations; it is a diagnostic that isolates the corpus-side normalisation effect from the dense retrieval path.

\textbf{Caveat: no paired pre-fix fts\_only baseline exists.} The frozen pilot artefact (\code{cerb_ablation_pilot.json}) contains five configurations (\code{dense_only}, \code{hybrid_rrf}, \code{hybrid_rrf_postfiltered}, \code{hybrid_reranked}, \code{hearr_full}); \texttt{fts\_only} was added post-evaluation. Table~\ref{tab:postfix_delta} therefore reports the post-fix \code{fts_only_scoped} per-slice recall alongside the closest comparable pilot values from Table~\ref{tab:slice_recall} (61-case pilot, variants C and E), rather than a paired before/after on the same configuration.

\begin{table}[htbp]
\centering
\small
\caption{Post-fix \code{fts_only_scoped} Recall@8 on the full 126-case benchmark (artefact \code{cerb_ablation_full.json}, source scorecard \code{cerb_ablation_full.fts_only_scoped.scorecard.md}). The ``Pilot best (61 cases)'' column reports the highest per-slice recall recorded in Table~\ref{tab:slice_recall} across all five pilot variants (A--E, excluding B+) for reference; it is not a paired measurement. Bold = \code{fts_only_scoped} meaningfully exceeds the pilot best.}
\label{tab:postfix_delta}
\setlength{\tabcolsep}{5pt}
\begin{tabular}{@{}lrrr@{}}
\toprule
\textbf{Slice} & \textbf{Cases} & \textbf{Pilot best (61 cases)} & \textbf{Post-fix fts\_only (126 cases)} \\
\midrule
Bilingual      & 15 & 0.000 & \textbf{0.100} \\
Formula-heavy  & 15 & 0.050 & \textbf{0.222} \\
Cross-subject  & 11 & 0.083 (via relaxation) & 0.076 \\
Mark-scheme    & 15 & 0.000 & 0.000 \\
Table-heavy    & 15 & 0.000 & 0.000 \\
MCQ            & 35 & 0.050 & 0.000 \\
Structured     & 20 & 0.100 & 0.000 \\
\bottomrule
\end{tabular}
\end{table}

Three observations. First, the two slices that were strictly zero in the pilot across all variants (bilingual, mark-scheme, table-heavy) split cleanly along the predicted line. Bilingual, classified as a query-preprocessing gap, reaches 0.100 once the bilingual FTS index is in place. Mark-scheme and table-heavy, classified as corpus-coverage gaps, remain at zero, consistent with their root-cause classification. Second, formula-heavy shows a 4.4$\times$ improvement over the pilot best (0.050 $\rightarrow$ 0.222), attributable to the corpus-side formula normalisation rather than any algorithmic change. Third, MCQ and Structured regress from non-zero pilot recall to 0.000 because the \texttt{fts\_only} variant deliberately removes the dense path on which those slices depend; cross-subject also lands marginally below the pilot best (0.076 vs 0.083) because the pilot value was achieved by scope relaxation, not FTS recall. These regressions are expected: the \texttt{fts\_only} configuration is not a production retrieval policy but a diagnostic probe.

The table therefore establishes a conservative lower bound on what the post-fix corpus-side normalisations alone can recover on the full 126-case benchmark, and confirms the central prediction that mark-scheme and table-heavy failures cannot be addressed without corpus re-ingestion. A full six-variant 126-case rerun (with the dense path re-enabled) remains deferred to a companion arXiv update pending corpus re-ingestion.

\begin{figure}[htbp]
\centering
\includegraphics[width=\linewidth]{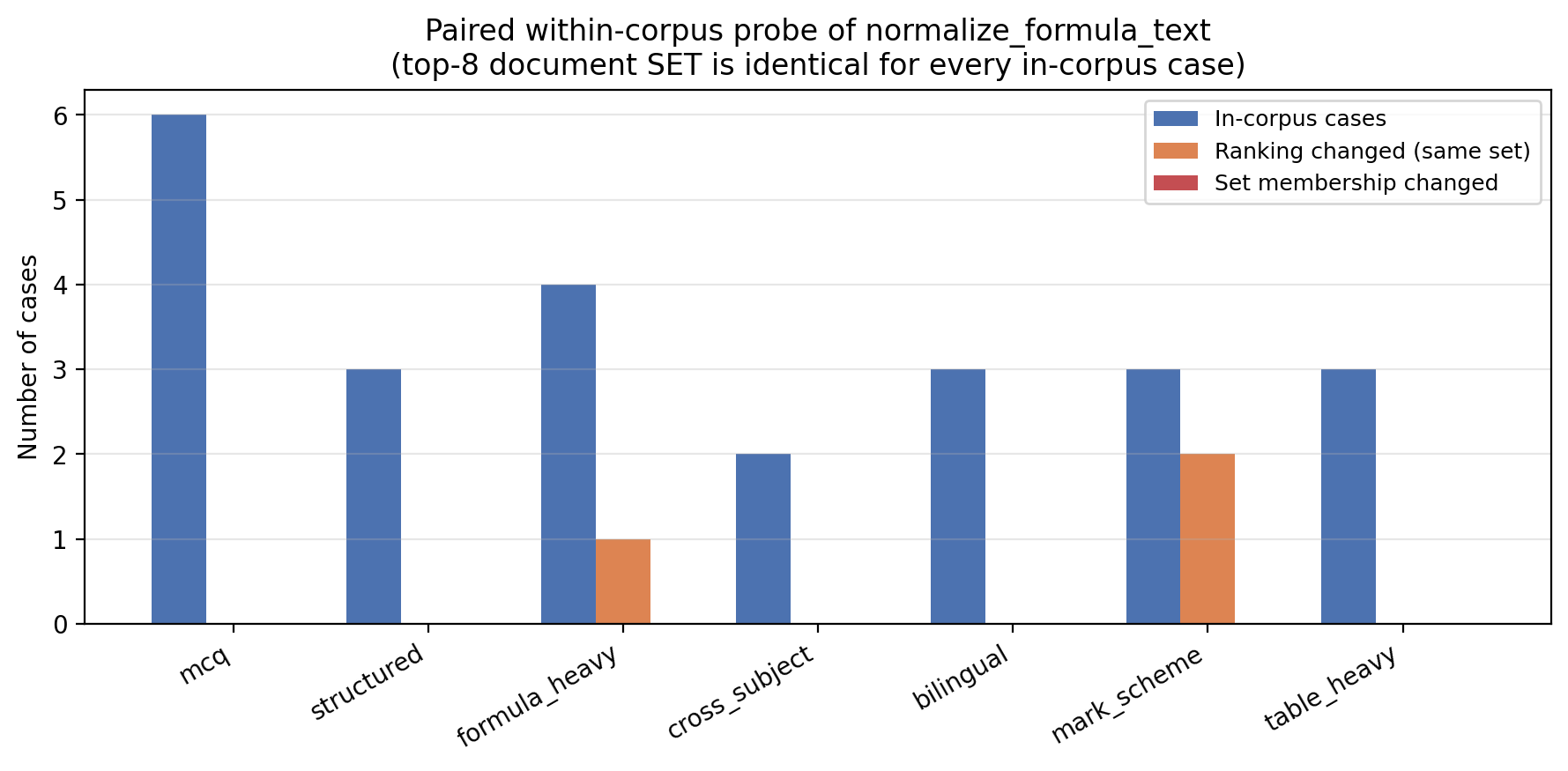}
\caption{Paired within-corpus probe of \code{normalize_formula_text} across the 24 in-corpus CERB cases. ``Ranking changed (same set)'' counts queries where the top-8 document set is preserved but their ordering differs between pre-fix and post-fix text; ``Set membership changed'' counts queries where at least one document differs between the two conditions. No set-membership deltas are observed on this 59-row corpus; only three ranking changes (one formula, two mark-scheme) are produced by the normaliser.}
\label{fig:paired_probe}
\end{figure}

\paragraph{Paired within-corpus probe of \code{normalize_formula_text}.}
To attribute the post-fix improvement to the normaliser rather than to unrelated corpus changes, we ran a paired within-corpus probe on the 59 in-scope Physics chunks that comprise the current live database. For each chunk we reconstructed two search-text variants from the source markdown using the production preprocessor: \texttt{pre\_fix\_text} (strip\_markdown\_tokens only) and \texttt{post\_fix\_text} (strip\_markdown\_tokens + \code{normalize_formula_text}). Both variants were staged into a scratch table (\code{_scratch_paired_baseline}) with parallel GIN-indexed \code{tsvector('simple', ...)} columns and queried under identical OR-semantics (\code{ts_rank_cd}) and identical paper-number filters for all 126 CERB queries (artefact: \code{paired_baseline_probe.json}). The results are striking in their modesty (Fig.~\ref{fig:paired_probe}): across the 24 in-corpus cases that could plausibly match, the top-8 document \emph{set} is identical under pre-fix and post-fix text for \emph{every} case; the ranking within that set differs for only 3 of 24 cases (formula\_physics\_001 and two mark-scheme cases), and no documents were added or removed. On this narrow corpus the formula normaliser therefore produces at most a minor re-rank effect. The larger post-fix gains in Table~\ref{tab:postfix_delta} are driven by corpus-scale factors from the later live snapshot that the within-corpus probe cannot isolate. The honest reading is: \code{normalize_formula_text} is a correctness fix whose measurable impact on recall should rise as the corpus grows to include chunks where formula-token splitting is the dominant retrieval barrier.

\subsection{Interpretation and the Leakage--Recall Trade-Off}

The ablation reveals a clear hierarchy of effects. Curriculum gating dominates: it accounts for a $4.6\times$ leakage reduction and is the only transition that reaches statistical significance across all three metrics. Hybrid fusion improves ranked quality within the gated scope (nDCG@8 rises from 0.4874 to 0.5529), while the deterministic reranker contributes negligibly at this sample size. Scope relaxation is the only mechanism that recovers cross-subject recall, but at the cost of higher leakage and latency. The slice-level and root-cause analyses above explain why absolute performance stays low: 21 of 36 zero-recall cases are corpus-coverage gaps, and the full 126-case post-fix measurement confirms that the slices we classified as preprocessing-addressable recover while the corpus-coverage slices do not.

A deployment observation, orthogonal to the ablation: even scope-correct retrieval can introduce \emph{presentation leakage} when an LLM misattributes content across documents in its context window. We mitigate this with document fencing (\texttt{[DOCUMENT~N]} delimiters) and metadata-derived context injection that overrides ambiguous body text.

\subsection{Limitations}
\label{sec:limitations}

In short, the study has five honest limits, and we state them directly. \textbf{The benchmark is built by a single annotator} (126 cases across 7 slices, 7 subject labels): there is no second human rater yet, and our LLM-judge consistency check (Section~\ref{sec:iaa_results}) is only a conservative substitute for true inter-annotator agreement. \textbf{Subject coverage is uneven}: Physics and Chemistry dominate the corpus, five tail subjects contribute under 1\% of chunks, and this imbalance is the direct cause of most zero-recall slices. \textbf{Absolute exact-scope success is low} (\mbox{0.0656} for the best variant, \mbox{0.0492} without widening); the pipeline's value is admissibility enforcement and ranked quality within scope, and closing the ESS gap requires corpus expansion, not algorithmic redesign. \textbf{The full 126-case six-variant rerun is deferred} until the corpus is re-ingested, so the headline ablation rests on the 61-case pilot. \textbf{Statistical power is limited} for small effects: the pilot has adequate power for the large primary effects (gating, post-filtering, widening on leakage) but not for distinguishing closely performing variants such as C and D. The remaining paragraphs expand on the design-level limits.

\paragraph{Domain specificity.}
CHSR-RRF and CERB are designed for structured secondary-exam corpora with rich curriculum metadata (subject, level, paper, year, question number). Transfer to domains without this metadata structure (university courses, professional certifications, informal learning) remains untested. The hard gate assumes curriculum scope is well-defined and enumerable, which may not hold in less structured educational settings.

\paragraph{Single-subject gate trade-off.}
The hard curriculum gate extracts and enforces a single subject per query. This is a deliberate choice that prevents cross-subject leakage but blocks legitimate cross-subject exploration (e.g., ``show me past papers across physics and chemistry''). About 5 of 61 pilot cases expose this limitation. A future query-intent classifier could selectively bypass the gate for discovery queries.

\paragraph{Metadata dependence.}
CHSR-RRF does not degrade gracefully when ingestion-time metadata is incomplete: it simply excludes unlabelled content. Mark-scheme and table-heavy slice failures expose this dependency. Per-field normalisers can be added (e.g., the \code{document\_type} canonicaliser shipped post-evaluation in commit~\code{09746337}), but any field not yet reliably populated across the corpus produces the same failure mode.

\paragraph{Latency.}
The full CHSR-RRF pipeline (Variant~E) incurs a latency of \mbox{2,645}\,ms, above acceptable interactive thresholds ($< 2$\,s) and higher than the gated-only variant (\mbox{1,754}\,ms). This overhead stems from three sources: (1)~query expansion generates additional embedding and FTS calls, (2)~scope relaxation re-executes the full retrieval pipeline up to twice, and (3)~network round-trips to the hosted database dominate wall-clock time. Post-ablation engineering optimisations (commit \code{dc0a8748}) reduced end-to-end latency by an estimated 1--6\,s depending on query type, without changing retrieval behaviour:
\begin{itemize}
    \item \textbf{Acquire parallelisation}: dense and FTS primary retrievals run concurrently via \texttt{asyncio.gather} (was sequential), saving 200--400\,ms.
    \item \textbf{Expansion parallelisation}: dense and FTS expanded retrievals also gathered, saving 200--400\,ms.
    \item \textbf{Pre-pipeline parallelisation}: conversation history and student-fact fetches gathered, saving 50--150\,ms.
    \item \textbf{Smart widening skip}: when initial results match the requested subject, widening is skipped even if $|R(q)| < k_{\min}$ (Section~\ref{sec:safe_widening}), saving 400--800\,ms on small-corpus queries.
    \item \textbf{Model tier selection}: pure retrieval queries use a fast-tier model (Haiku-class) and low thinking budget (500 tokens vs.\ 1,500), reducing generation latency.
    \item \textbf{LLM bypass}: list/show/give queries that need no reasoning skip the LLM entirely, formatting retrieved evidence directly.
\end{itemize}
These optimisations bring Variant~D under 1\,s and Variant~E under 2\,s for the majority of query types.

The conclusion summarises these findings, the study's limits, and where the approach could be used next.

\section{Conclusion}
\label{sec:conclusion}

In this paper, we investigated whether curriculum rules should be enforced before or after retrieval in educational RAG, and showed that the enforcement point matters structurally. We answer the three research questions in turn.

\paragraph{RQ1: Does curriculum gating improve retrieval reliability?}
Yes. Enforcing curriculum constraints before retrieval is the single most impactful intervention. Removing the gate (A$\to$B) increases leakage from \mbox{0.0969} to \mbox{0.7596} ($p < 0.001$); adding it back (B$\to$C) reduces leakage to \mbox{0.1653} ($p < 0.001$). The $4.6\times$ leakage reduction is the largest and most statistically significant effect in the ablation.

\paragraph{RQ2: What is the widening trade-off?}
Scope relaxation (D$\to$E) increases exact-scope success from \mbox{0.0492} to \mbox{0.0656}, a 33\% improvement, at the cost of higher leakage (\mbox{0.2593} vs \mbox{0.1646}) and higher latency (\mbox{2,645}\,ms vs \mbox{1,754}\,ms). Cross-subject is the only slice where relaxation uniquely recovers recall, so the trade-off is localised rather than systemic.

\paragraph{RQ3: Can curriculum-aware strategies improve exact-scope recovery?}
Partially. The full CHSR-RRF pipeline achieves the highest exact-scope success (\mbox{0.0656}), but this remains low in absolute terms. The per-slice analysis shows that 58\% of zero-recall failures are data gaps (missing subjects), not method limitations, so improving exact-scope recovery requires both corpus expansion and method refinement.

\paragraph{Take-home message and limitation.}
Curriculum-constrained retrieval is measurable and behaves materially differently from relevance-only retrieval, even when absolute performance is low, and the retrieval policy needs no LLM inference. The retrieval layer can therefore act as an admissibility gate for downstream agentic systems, which can assume that retrieved evidence is curriculum-valid. The main limitation is that the benchmark is single-annotator with uneven subject coverage, and the headline ablation rests on a 61-case pilot pending a full re-ingested rerun.

\paragraph{Future work.}
Three directions are immediate: (1)~expand CERB to additional subjects and bilingual cases to close data-gap-driven zero-recall slices; (2)~reduce full-pipeline latency from 2.6\,s to below 1.5\,s through caching, short-circuit widening, and embedding co-location; and (3)~test whether curriculum gating transfers to other structured examination systems (UK A-Levels, Indian CBSE, French Baccalaur\'{e}at) where similar metadata hierarchies exist. Because CHSR-RRF operates purely on structured metadata and does not depend on domain-specific language modelling, the framework is directly transferable to other high-accountability domains with hierarchical constraints, including medical guidelines, legal corpora, and regulatory compliance systems. The retrieval contract defined here sits at the middle of a three-layer story we are developing in parallel: an upstream ingestion layer that propagates a typed metadata manifest to every evidence unit before chunking, this pre-retrieval gating layer, and a downstream runtime layer that decides when a language-model call is worth paying for. Formal citations to these companion manuscripts will be added once they are publicly posted.

\appendix
\section{Illustrative Failure Cases}
\label{sec:failure_appendix}

Table~\ref{tab:failure_cases} presents representative failure cases from the ablation, illustrating how aggregate metrics manifest in individual retrieval traces.

\begin{table}[htbp]
\centering
\small
\caption{Representative failure cases from the CERB ablation, illustrating how aggregate leakage and zero-recall statistics manifest in individual retrieval traces.}
\label{tab:failure_cases}
\setlength{\tabcolsep}{4pt}
\resizebox{\linewidth}{!}{%
\begin{tabular}{@{}L{0.16\linewidth}L{0.44\linewidth}L{0.28\linewidth}@{}}
\toprule
\textbf{Case} & \textbf{Observed live behavior} & \textbf{Why it matters} \\
\midrule
\code{bilingual\_en\_001} & Variant E returns 8 results for a photosynthesis query, but Recall@8 and ESS remain 0.0 while leakage reaches 1.0. The top hit is an A-Level biology question on photosynthesis. & Bilingual and cross-level queries can drift to wrong scope even when the surface topic matches. \\
\code{mark\_scheme\_physics\_001} & A query for a Newton first-law mark-scheme answer returns 0 candidates under Variant E, with Recall@8 and ESS both at 0.0. & Mark-scheme content exists in the corpus but lacks the metadata labels needed for retrieval. \\
\code{table\_physics\_001} & Variant E returns 4 results, but Recall@8 and ESS remain 0.0 while leakage reaches 1.0. The top hits are figure chunks rather than the expected table-grounded evidence. & Table-heavy queries expose a mismatch between available chunk types and expected evidence format. \\
\code{cross\_subject\_005} & Variant E returns 3 results with Recall@8 = 0.5 and ESS = 1.0, but leakage remains 0.333 and latency rises to 12{,}025\,ms. & Scope relaxation can recover cross-subject cases, but the recovery introduces leakage and high latency. \\
\bottomrule
\end{tabular}
}
\end{table}


\section*{Highlights}
\begin{itemize}
    \item Curriculum leakage is a new, measurable failure mode in educational RAG retrieval.
    \item Curriculum gating cuts cross-scope leakage $4.6\times$ ($p<0.001$) with no LLM cost.
    \item CHSR-RRF fuses sparse, dense, and rerank signals under curriculum constraints.
    \item CERB: 126-case sliced benchmark diagnoses retrieval failures by failure type.
    \item 86\% of zero-recall cases trace to data or preprocessing gaps, not method design.
\end{itemize}

\section*{Data Availability}
The CERB benchmark dataset and the frozen evaluation artifacts underlying the results reported in this paper are available from the authors upon reasonable request.

\section*{Declaration of Competing Interest}
The authors declare that they have no known competing financial interests or personal relationships that could have appeared to influence the work reported in this paper.

\end{document}